\documentclass[aps,prb,reprint,showpacs,superscriptaddress,footinbib]{revtex4-2}
\usepackage{graphicx}
\usepackage{amsmath}
\usepackage{amsthm}
\usepackage{amssymb}
\usepackage{dcolumn}
\usepackage{bm}
\usepackage{color}
\usepackage[utf8]{inputenc}
\usepackage[T1]{fontenc}
\usepackage{booktabs, array, mathptmx, float, tabularx, multirow}
\usepackage{siunitx, xcolor}
\usepackage{epstopdf}
\usepackage{grffile}
\usepackage{indentfirst}
\usepackage{mathrsfs}
\usepackage{comment}
\usepackage{lmodern}
\usepackage[colorlinks,linkcolor=blue,anchorcolor=blue,citecolor=blue,urlcolor=blue,bookmarks=false]{hyperref}

\begin{document}

\title{Differentiable Inverse Design of Two-Body Interactions for the Moore--Read State}

\author{Yi Yang}
\affiliation{School of Physics, Hangzhou Normal University, Hangzhou 311121, People's Republic of China}
\affiliation{Department of Physics and Chongqing Key Laboratory for Strongly Coupled Physics, Chongqing University, Chongqing 401331, People's Republic of China}
\affiliation{Zhejiang Institute of Modern Physics and Zhejiang Key Laboratory of Micro-nano Quantum Chips and Quantum Control, Zhejiang University, Hangzhou 310027, People's Republic of China}

\author{Xin Wan \thanks{Corresponding author. Email: xinwan@zju.edu.cn}}
\email{xinwan@zju.edu.cn}
\affiliation{Zhejiang Institute of Modern Physics and Zhejiang Key Laboratory of Micro-nano Quantum Chips and Quantum Control, Zhejiang University, Hangzhou 310027, People's Republic of China}

\author{Zi-Xiang Hu \thanks{Corresponding author. Email: zxhu@cqu.edu.cn}}
\email{zxhu@cqu.edu.cn}
\affiliation{Department of Physics and Chongqing Key Laboratory for Strongly Coupled Physics, Chongqing University, Chongqing 401331, People's Republic of China}

\begin{abstract}
   Designing interactions that stabilize target quantum states is a central challenge in strongly correlated physics. For fractional quantum Hall states, identifying suitable two-body interactions remains difficult even when a target wave function is known. Here we implement a differentiable inverse-design approach that combines automatic differentiation with exact diagonalization to optimize two-body interactions, using the Moore--Read Pfaffian state as a benchmark. The optimized ground states achieve overlaps exceeding 99\% with the target for systems containing up to 12 electrons. Excitation and entanglement spectra provide complementary finite-size diagnostics. A common exponential pseudopotential profile retains an overlap of 91.1\% at 14 electrons without further optimization, supporting its transferability beyond the optimization sizes. These results demonstrate the approach on a non-Abelian quantum Hall benchmark and motivate its application to other numerically accessible trial states for which suitable approximate parent Hamiltonians remain unknown.
\end{abstract}

\date{\today}

\maketitle

\section*{Introduction}
The fractional quantum Hall effect (FQHE) provides a paradigmatic platform for exploring strongly correlated topological phases of matter. The even-denominator state at $\nu=5/2$ is of particular interest as a candidate for non-Abelian topological order, with potential applications in fault-tolerant topological quantum computation~\cite{FQHE,even,TQC1,TQC2,TQC3,TQC4,TQC5,TQC6,TQC7,TQC8,TQC9,TQC10,TQC11,TQC12,TQC13,TQC14}. Identifying the topological order realized in this state remains an important theoretical and experimental question. Prominent candidate states include the Moore--Read (MR) Pfaffian, its particle--hole conjugate (anti-Pfaffian), and the particle--hole-symmetric Pfaffian (PH-Pfaffian)~\cite{mooreread,MR1,MR2,MR3,MR4,MR5,antiPfaffian,PHPfaffian1,PHPfaffian2,V1V33}. The MR Pfaffian is the exact zero-energy ground state of a short-range three-body parent Hamiltonian. In real electronic systems, however, the dominant interaction is the two-body Coulomb repulsion projected into a partially filled Landau level. This raises a fundamental question: Can two-body interactions that closely approximate a prescribed fractional quantum Hall state be identified systematically?

Most theoretical approaches in condensed matter physics address the forward problem: predicting emergent phases from a prescribed Hamiltonian. Here we instead consider the inverse problem: how should microscopic interactions be designed to realize a desired quantum phase? Recent advances in automatic differentiation and differentiable programming have made this inverse-design paradigm computationally feasible for strongly correlated quantum systems~\cite{inverse1,inverse2,inverse3,inverse4,inverse5,inverse6}. 

Several physical mechanisms, including Landau-level mixing, finite quantum-well width, dielectric screening, and tilted magnetic fields, can modify effective electron-electron interactions and modulate the stability of the Pfaffian phase~\cite{PhysRevLett.105.096802,PhysRevLett.106.116801,Smith_2012,PhysRevB.87.155426,PhysRevB.87.245425,PhysRevX.5.021004,YLPCPB,ML2,LLMka,thickness1,thickness2,PhysRevBHu19}. Nevertheless, these approaches explore only physically constrained directions in interaction space rather than systematically searching for interactions optimized toward a prescribed topological phase. Meanwhile, rapid progress in quantum simulation has created unprecedented opportunities to engineer effective interactions beyond those available in conventional semiconductor systems~\cite{QS1,QS2,QS3,QS4,QS5,QS6,QS7,QS8,QS9,QS10}. Together with recent advances in differentiable programming and gradient-based optimization~\cite{ML1,ML2,ML3,ML4,ML5,ML6,ML7,ML8,ML9,ML10}, these developments motivate a more systematic framework for inverse Hamiltonian design.

Previous studies have explored the relation between two-body interactions and Moore--Read physics. Rezayi et al.~\cite{MR2} investigated the competition between paired and competing phases, while Storni et al.~\cite{Storni2010} examined the dependence of the Pfaffian overlap and excitation gap on $V_1$ and $V_3$. Hutzel et al.~\cite{V1V33} studied a particle--hole-symmetric two-body model obtained by symmetrizing the three-body Pfaffian Hamiltonian. More generally, Greiter et al.~\cite{Greiter2018} developed a method for identifying exact and approximate parent Hamiltonians from trial states, and Pakrouski~\cite{inverse3} formulated an optimization-based Hamiltonian-design approach, including two-body approximations to the Moore--Read state and tests of transferability to larger systems. Our work builds on these developments through an implementation that differentiates the ground-state overlap through exact diagonalization using automatic differentiation, whereas the numerical optimization in Ref.~\cite{inverse3} employed finite-difference gradients. 

We use the Moore--Read Pfaffian as a benchmark for differentiable inverse Hamiltonian design. Optimizing two-body Haldane pseudopotentials yields ground-state overlaps exceeding $99\%$ for systems with 6, 8, 10, and 12 electrons, while excitation and entanglement spectra provide complementary finite-size diagnostics. A common exponential fit to the optimized pseudopotentials retains a Pfaffian overlap of $91.1\%$ for 14 electrons without further optimization, demonstrating transferability beyond the optimization sizes within the accessible finite-size range. These results illustrate how automatic differentiation through exact diagonalization can guide the search for two-body interactions approximating a prescribed target state, and motivate applications to other numerically accessible trial states whose suitable approximate parent Hamiltonians remain unknown.

\section*{Model and methods}\label{sec:methods}

\begin{figure*}[t]
\centering
\includegraphics[width=\textwidth]{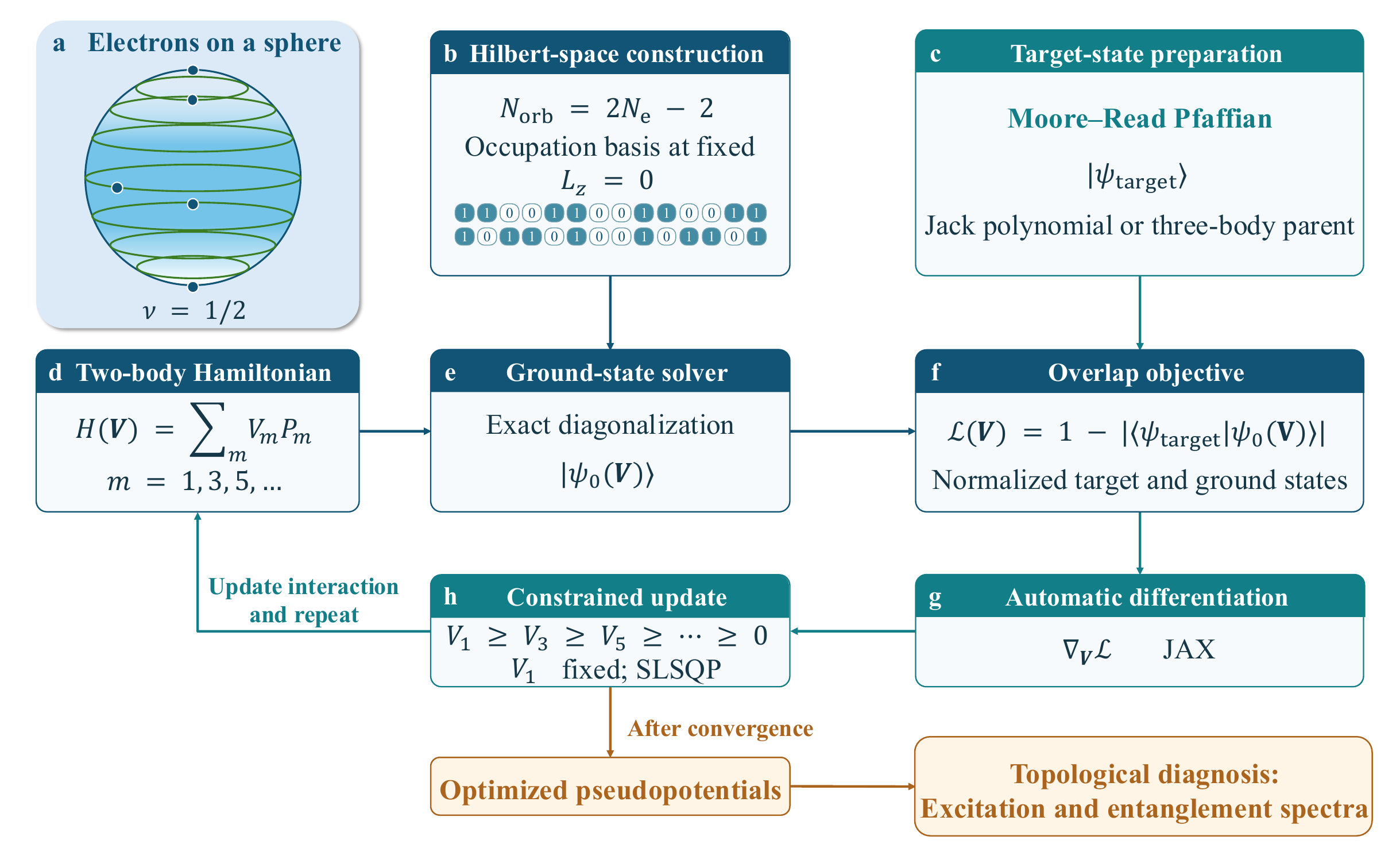}
\par\smallskip
{\small Fig.~\ref{model}\par}
\end{figure*}

\subsection*{Spherical geometry and interaction Hamiltonian}

We formulate inverse Hamiltonian design by parameterizing a rotationally invariant two-body interaction in terms of Haldane pseudopotentials on Haldane's spherical geometry~\cite{sphere1,sphere2}. We consider fully spin-polarized electrons confined to the surface of a sphere threaded by a magnetic monopole of strength $Q$. After projection into a single Landau level, the interaction Hamiltonian is written as
\begin{equation}
    \hat{H}_{\mathrm{int}} = \sum_{i<j}\sum_m V_m \hat{P}_{ij}^{(m)},
\end{equation}
where $\hat{P}_{ij}^{(m)}$ projects the pair $(i,j)$ onto the relative-angular-momentum channel $m$. For fully spin-polarized fermions, only odd values of $m$ contribute due to the Pauli principle. We choose the orbital number according to the Moore--Read Pfaffian counting relation, $N_{\mathrm{orb}}=2N_e-2$, corresponding to single-particle angular momentum $l=(N_{\mathrm{orb}}-1)/2$. For the Coulomb reference in Landau level $n$, the physical monopole strength is $Q=l-n$, so that $N_{\mathrm{orb}}=2(Q+n)+1$. The overall energy scale is fixed by setting $V_1$ to the spherical 0LL Coulomb value, while the remaining odd pseudopotentials $\{V_m\}_{m\ge3}$ are optimized. For $N_e=12$ with 22 orbitals, the fixed value is $V_1=0.4681236240\,e^2/(\epsilon\ell_B)$, where $\ell_B$ is the magnetic length and $\epsilon$ is the dielectric constant. The 1LL Coulomb reference retains its own pseudopotentials, with $V_1=0.4716484651\,e^2/(\epsilon\ell_B)$.

\subsection*{Target state and differentiable optimization}

The optimization target is chosen to be the Moore--Read Pfaffian state,
\begin{equation}
    \Psi_{\mathrm{MR}}(\{u_i, v_i\}) = \operatorname{Pf}\left( \frac{1}{u_i v_j - u_j v_i} \right) \prod_{i<j} (u_i v_j - u_j v_i)^2,
\end{equation}
where $(u,v)$ are the spinor coordinates on the sphere. This wave function can be obtained either as the exact zero-energy ground state of the three-body Pfaffian parent Hamiltonian~\cite{mooreread} or via the Jack polynomial construction~\cite{jack1,jack2}. Given a specified set of pseudopotentials, we diagonalize the two-body Hamiltonian and denote the resulting ground state by $|\Psi_0(\{V_m\})\rangle$. The inverse-design problem is formulated as minimizing the overlap-based objective
\begin{equation}
\mathcal{L}(\{V_m\}) =
1-\left|\langle \Psi_{\mathrm{MR}}|\Psi_0(\{V_m\})\rangle\right| .
\end{equation}
In finite-size exact diagonalization, the overlap directly quantifies the similarity between the optimized ground state and the target wave function. Spectral and entanglement diagnostics are used below to further characterize the optimized states. Automatic differentiation is implemented within a JAX-enabled exact-diagonalization framework to compute gradients of the ground-state overlap with respect to the pseudopotentials without finite-difference approximations~\cite{jax}. For an isolated, nondegenerate ground state, the overlap objective is differentiable wherever the overlap is nonzero. We restrict the search to nonnegative, monotonically decreasing pseudopotential sequences by imposing $0\le V_m\le V_1$ and $V_1\ge V_3\ge V_5\ge\cdots\ge0$. These constraints define the interaction family explored here; nonnegativity and monotonicity are imposed a priori rather than inferred from the optimization. No exponential functional form is imposed. Constrained optimization is performed using the SLSQP algorithm, with gradients supplied by automatic differentiation. To reduce sensitivity to local minima, we employ a multi-start strategy in which the optimization is initialized from many random pseudopotential profiles. We retain the solution with the highest ground-state overlap among the runs. 

For the $N_e=12$ benchmark, automatic differentiation reduces the optimization wall-clock time by factors of $2.72$ and $3.63$ relative to forward finite differences with steps of $10^{-6}$ and $10^{-7}$, respectively, under identical initialization, constraints, and stopping criteria (Supplementary Note 7 and Supplementary Table 2).

\subsection*{Spectral and entanglement diagnostics}

To characterize the optimized states beyond wave-function overlap, we examine two complementary finite-size diagnostics. First, the neutral excitation spectrum is computed by diagonalizing the full Hamiltonian, and each eigenstate is assigned a total angular momentum $L$. This labeling allows us to identify the unique $L=0$ ground state and to track the lowest neutral excitations as a function of $L$. Second, the orbital entanglement spectrum (OES) is evaluated from a bipartition of contiguous orbitals on the sphere~\cite{entanglement1}. For the Moore--Read phase, the low-lying OES levels exhibit the characteristic edge-mode counting of the associated conformal field theory, with different sequences in even and odd particle-number sectors~\cite{entanglement2,entanglement3}. Together, these diagnostics test whether the optimized finite-size states exhibit spectral and entanglement features consistent with the Moore--Read state. The overall inverse-design workflow is summarized in Fig.~\ref{model}. 

\section*{Results}\label{sec:results}
\subsection*{Optimized interactions and ground-state properties}

\subsubsection*{Inverse-designed interaction profile}

We begin by examining the interaction profile identified through differentiable inverse Hamiltonian design. Figure~\ref{fig:pps12} shows the optimized Haldane pseudopotentials together with the Coulomb pseudopotentials projected into the lowest (0LL) and first excited (1LL) Landau levels. The optimized interaction remains monotonically decreasing but exhibits a markedly shorter-ranged profile than either Coulomb interaction. Long-range pseudopotential components are strongly suppressed, whereas the leading short-range channels remain comparable to those of the Coulomb interactions. The optimized value of $V_3$ lies between those of the 0LL and 1LL Coulomb interactions, highlighting the importance of the leading short-range pseudopotential channels for stabilizing the Pfaffian phase. This observation is broadly consistent with previous studies showing that appropriate tuning of the short-range pseudopotentials can substantially enhance the Pfaffian overlap~\cite{V1V33}. The optimized Hamiltonian achieves an overlap of $99.49\%$ with the Moore--Read Pfaffian state, demonstrating that the inverse-designed two-body interaction faithfully reproduces the target many-body wave function within the accessible finite-size Hilbert space. We next characterize this state through its excitation and entanglement spectra.

\begin{figure}[t]
\centering
\includegraphics[width=\columnwidth]{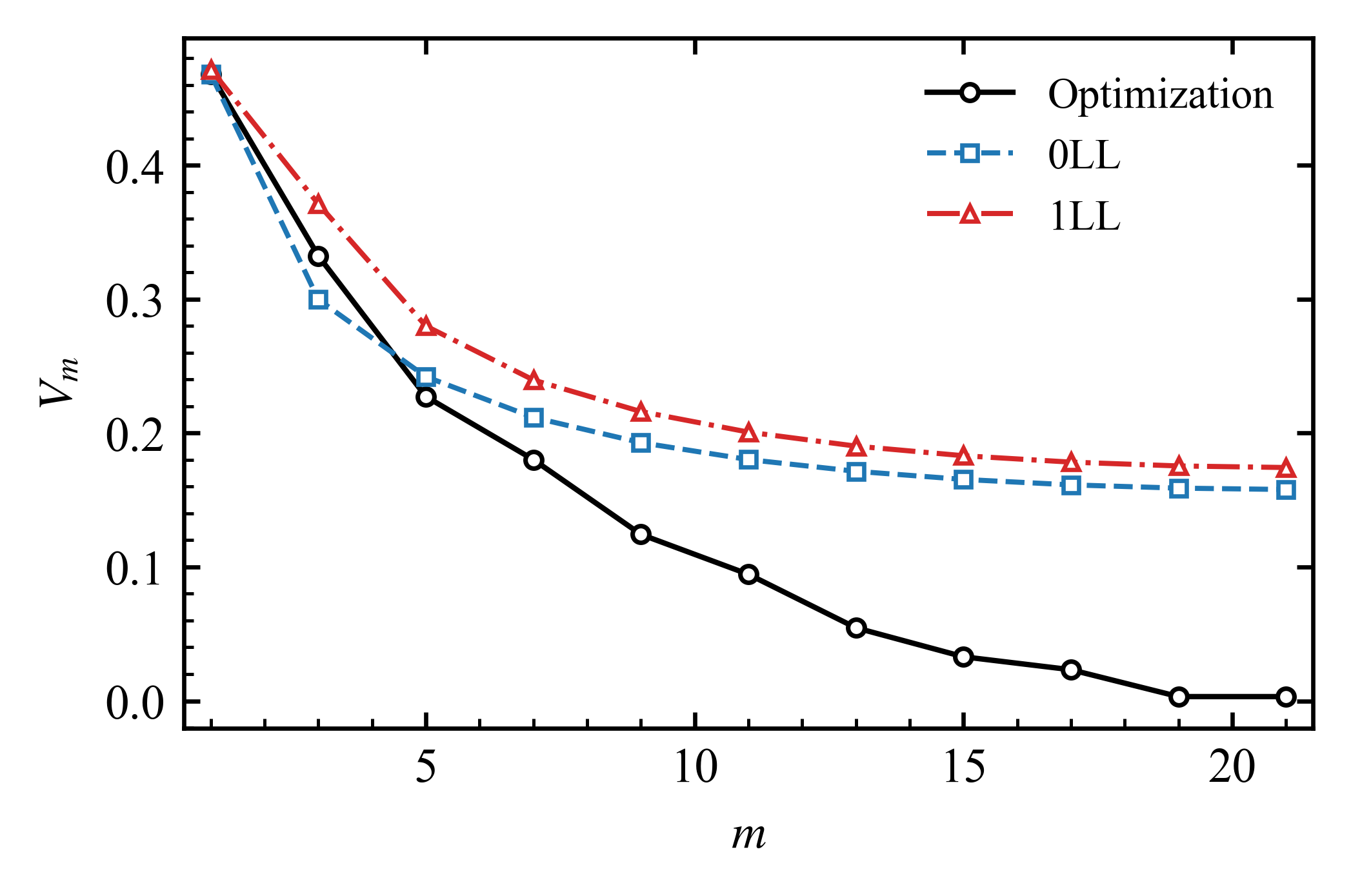}
\par\smallskip
{\small Fig.~\ref{fig:pps12}\par}
\end{figure}

\subsubsection*{Energy spectrum}

To characterize the low-energy structure, we examine the neutral excitation spectrum at the Moore--Read shift, as shown in Fig.~\ref{fig:energy}. Both the optimized interaction and the Coulomb interaction projected into the first excited Landau level yield an isolated $L=0$ ground state. However, their excitation spectra display pronounced differences. The optimized interaction produces a more clearly separated low-energy collective branch, accompanied by larger excitation energies over a broad range of momenta. As a result, the neutral excitation gap is enhanced compared with the 1LL Coulomb interaction. Thus, the optimized interaction combines high Pfaffian overlap with a larger neutral excitation gap in this finite system.

\begin{figure}[t]
\centering
\includegraphics[width=\columnwidth]{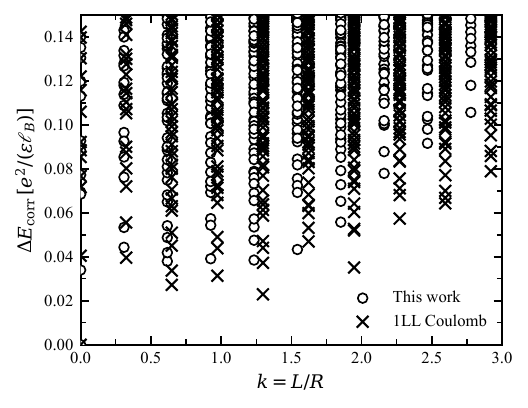}
\par\smallskip
{\small Fig.~\ref{fig:energy}\par}
\end{figure}

\subsubsection*{Orbital entanglement spectrum}

We further characterize the optimized ground state through the OES, obtained from contiguous orbital partitions on the sphere. Figure~\ref{fig:OES} compares the OES of the optimized state with that of the exact Pfaffian state for two representative cuts, $N_a=6$ and $N_a=7$. In both cases, the low-lying entanglement spectrum of the optimized state reproduces the characteristic Pfaffian counting within each $L_z^A$ sector~\cite{entanglement1}. Specifically, the level counting follows $(1,1,3,5,\ldots)$ for $N_a=6$ and $(1,2,4,7,\ldots)$ for $N_a=7$, consistent with the expected edge excitation spectrum of the Moore--Read phase~\cite{mooreread}. The low-lying levels also closely match the dispersion of the Pfaffian entanglement spectrum and are separated from higher nonuniversal levels by a well-defined entanglement gap. These finite-size diagnostics are consistent with the characteristic entanglement structure of the Moore--Read Pfaffian state. Additional entanglement-entropy results are presented in Supplementary Note 1 and Supplementary Figure 1.
\begin{figure}[t]
\centering
\includegraphics[width=\columnwidth]{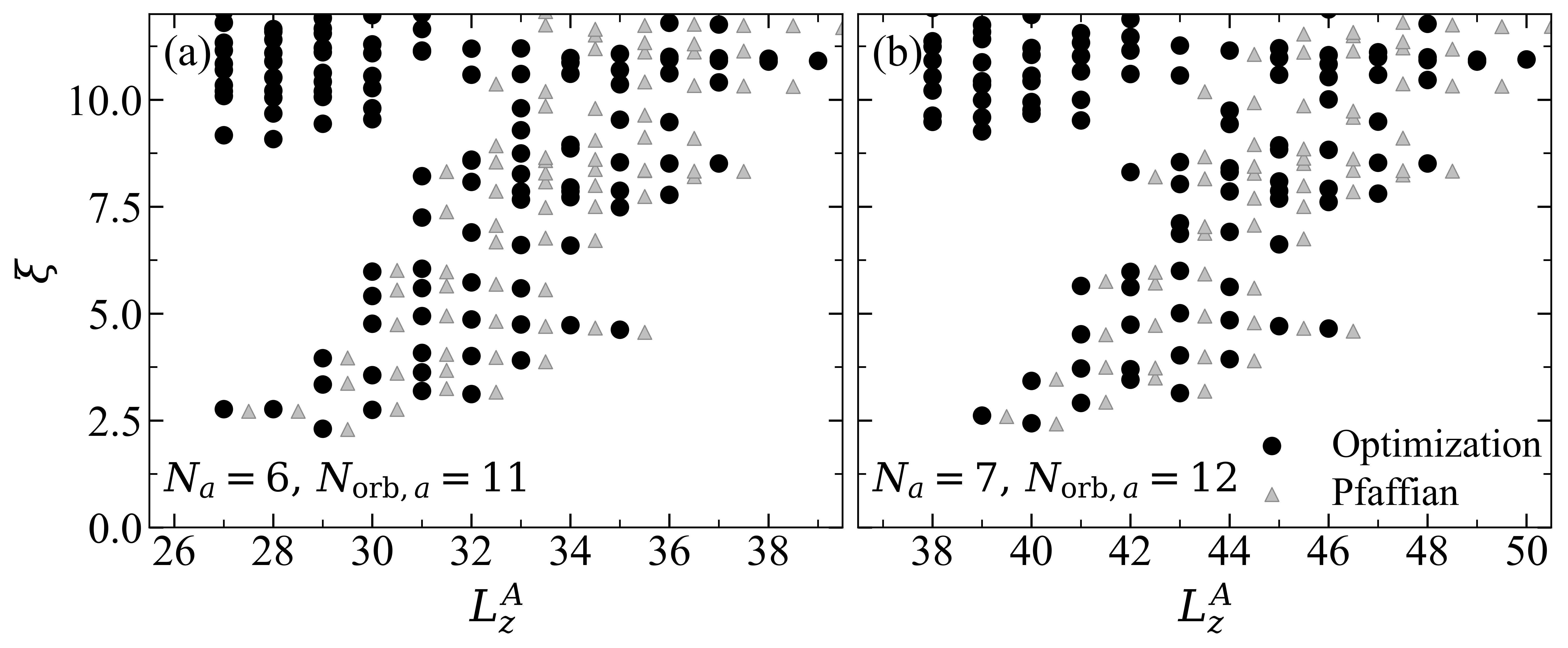}
\par\smallskip
{\small Fig.~\ref{fig:OES}\par}
\end{figure}

We next compare the optimized interaction with the physically relevant Coulomb interaction projected into the first excited Landau level. For $N_e=12$, the 1LL Coulomb ground state has an overlap of $0.819$ with the MR Pfaffian state, consistent with previous finite-size studies showing that the 1LL Coulomb interaction lies close to the Pfaffian phase. Nevertheless, the optimized interaction achieves a substantially larger overlap and produces a more clearly separated neutral excitation branch. This improvement is further reflected in the OES shown in Fig.~\ref{fig:comparison1LLOES}. While the 1LL Coulomb state exhibits stronger mixing between Pfaffian-like low-lying levels and higher entanglement branches, the optimized state displays a more clearly resolved universal low-lying structure and a larger entanglement gap. These results indicate that although the Coulomb interaction is close to the Pfaffian regime, it does not correspond to the optimal two-body interaction within the explored pseudopotential space.

\begin{figure}[t]
\centering
\includegraphics[width=\columnwidth]{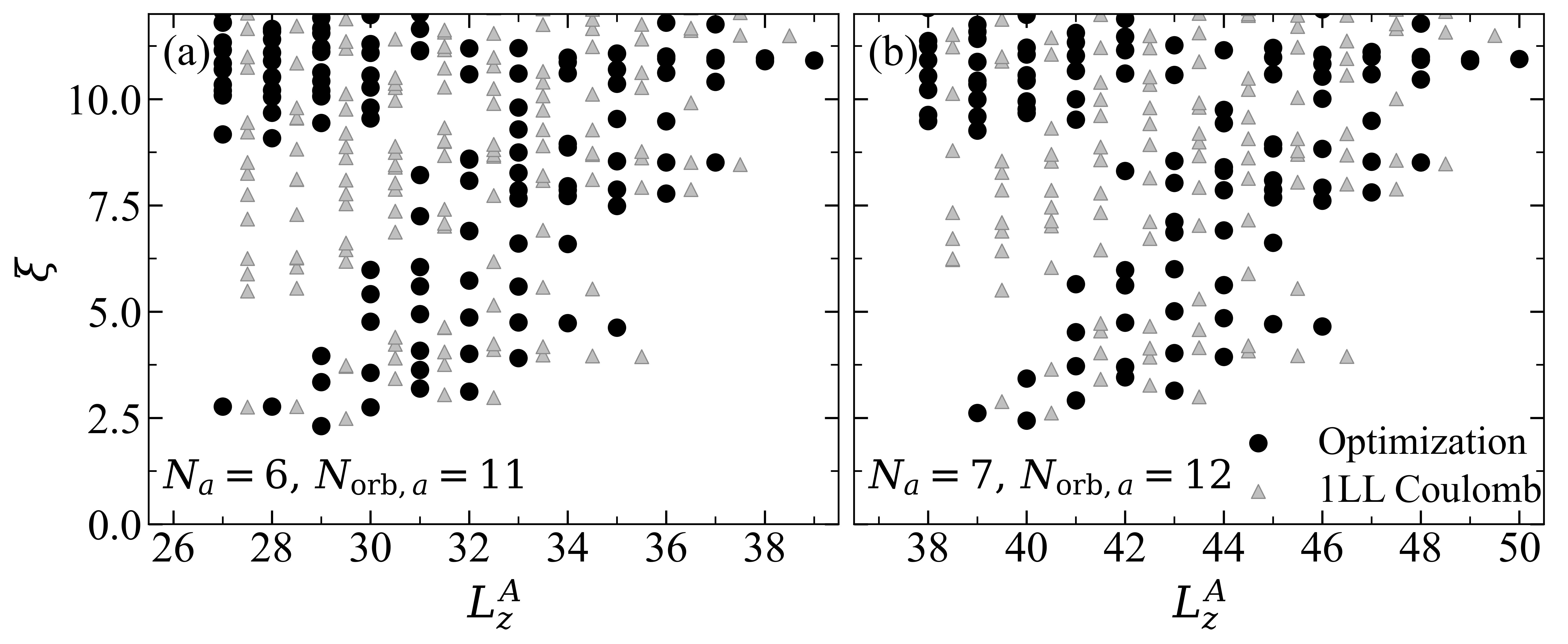}
\par\smallskip
{\small Fig.~\ref{fig:comparison1LLOES}\par}
\end{figure}

Together, these results show that the optimized two-body interaction closely approximates the Pfaffian target and reproduces its characteristic finite-size spectral and entanglement features. The Coulomb interaction lies close to, but does not coincide with, the pseudopotential configuration that maximizes the Pfaffian overlap within the restricted space of two-body interactions considered here.

\subsection*{Finite-size dependence and transferability}
\label{sec:larger}

\begin{figure}[ht]
\centering
\includegraphics[width=\columnwidth]{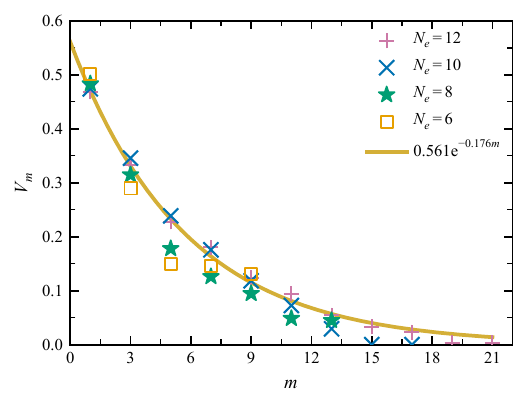}
\par\smallskip
{\small Fig.~\ref{fig:fitpps}\par}
\end{figure}

We next investigate the finite-size dependence of the optimized interaction within the constrained pseudopotential family. Figure~\ref{fig:fitpps} shows the independently optimized odd pseudopotentials for $N_e=6,8,10$, and $12$ at the MR shift. Monotonicity is imposed in each optimization, whereas the approximate agreement between the profiles at different sizes and their description by a common exponential form are empirical observations within the accessible size range. We fit these profiles using
\begin{equation}
\label{eqfit}
V_m=Ae^{-\alpha m},
\end{equation}
with a global fit corresponding to $A\simeq0.561e^2/(\epsilon\ell_B)$ and $\alpha\simeq0.176$, where $\alpha$ is the dimensionless decay coefficient. Overlaps obtained from the fitted interaction and the Coulomb reference interactions are summarized in Supplementary Note 2 and Supplementary Table 1.

The compact exponential description does not imply that only the leading channels are needed. Directly setting $V_m=0$ for $m>7$ in the independently optimized interactions, without changing the retained coefficients or reoptimizing, reduces the overlap from $0.99976143$ to below $10^{-12}$ for $N_e=10$ and from $0.99494328$ to $0.06030019$ for $N_e=12$. The ground-state angular momentum changes from $L=0$ to $L=2$ in the former case and remains $L=0$ in the latter. These tests show that the smaller higher-$m$ coefficients contribute materially to the optimized states; they do not determine the best overlap attainable after reoptimizing only the retained channels.

\subsubsection*{Finite-size dependence of the neutral gap}
To examine a single interaction across different system sizes, we fix the odd pseudopotentials to $V_m=0.561e^{-0.176m}$ in units of $e^2/(\epsilon\ell_B)$, without further optimization or size-dependent rescaling. Figure~\ref{fig:gapoverlap} compares the density-corrected neutral gaps and MR overlaps of this fitted interaction with those of the spherical 1LL Coulomb interaction for $N_e=8,10,12$, and $14$. The neutral gap is defined as the energy difference between the ground state and the lowest excited state at fixed particle number and orbital number. All ground states shown have $L=0$. We apply the density correction $\Delta E_{\mathrm{corr}}=\sqrt{2Q\nu/N_e}(E_1-E_0)$ with $\nu=1/2$, using $Q=(N_{\mathrm{orb}}-1)/2$ for the fitted interaction represented in the 0LL and $Q=(N_{\mathrm{orb}}-3)/2$ for the 1LL Coulomb reference.

Under these conventions, the fitted interaction has a larger density-corrected neutral gap than the 1LL Coulomb reference at each displayed size, although the difference is small for $N_e=12$ and $14$. Its MR overlap remains above $0.89$ over this range. The comparison focuses on $N_e=8$--$14$; at $N_e=6$, the fixed fitted interaction places the high-overlap MR-like state in the first excited level rather than the ground state.

\begin{figure}[t]
\centering
\includegraphics[width=\columnwidth]{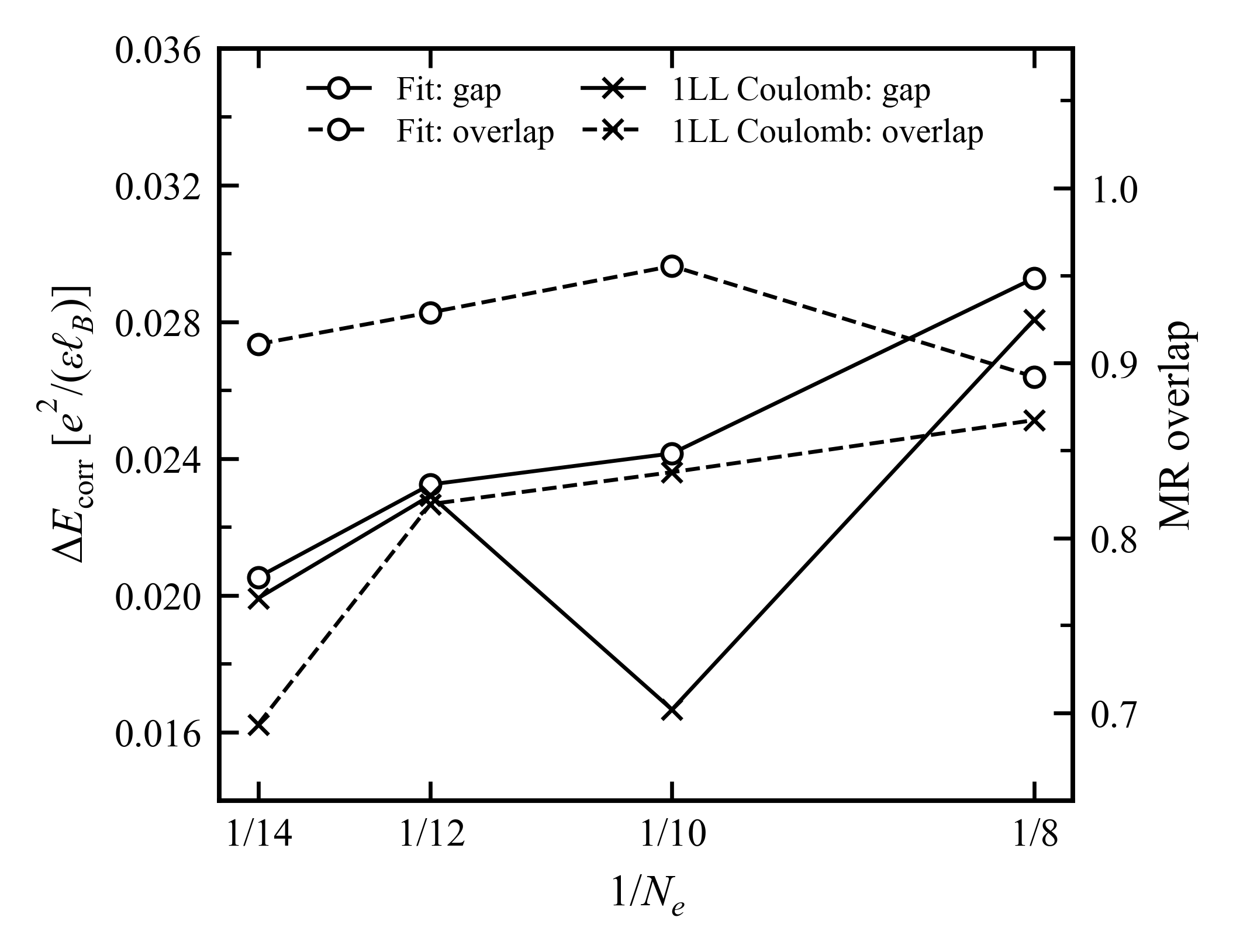}
\par\smallskip
{\small Fig.~\ref{fig:gapoverlap}\par}
\end{figure}

\subsubsection*{Transfer to a larger system}
To examine the transferability of the optimized interaction, we construct the $N_e=14$ Hamiltonian using the fitted exponential pseudopotential profile in Eq.~(\ref{eqfit}), without additional optimization. The resulting ground state has an overlap of $91.1\%$ with the MR Pfaffian state. Although this value is lower than the $>99\%$ overlaps obtained from direct optimization at smaller system sizes, it remains substantial given that the interaction parameters are fixed entirely from the finite-size trend. This result indicates that the fitted pseudopotential profile captures the dominant interaction features underlying the Pfaffian correlations beyond the systems used in the optimization.
\begin{figure}[ht]
\centering
\includegraphics[width=\columnwidth]{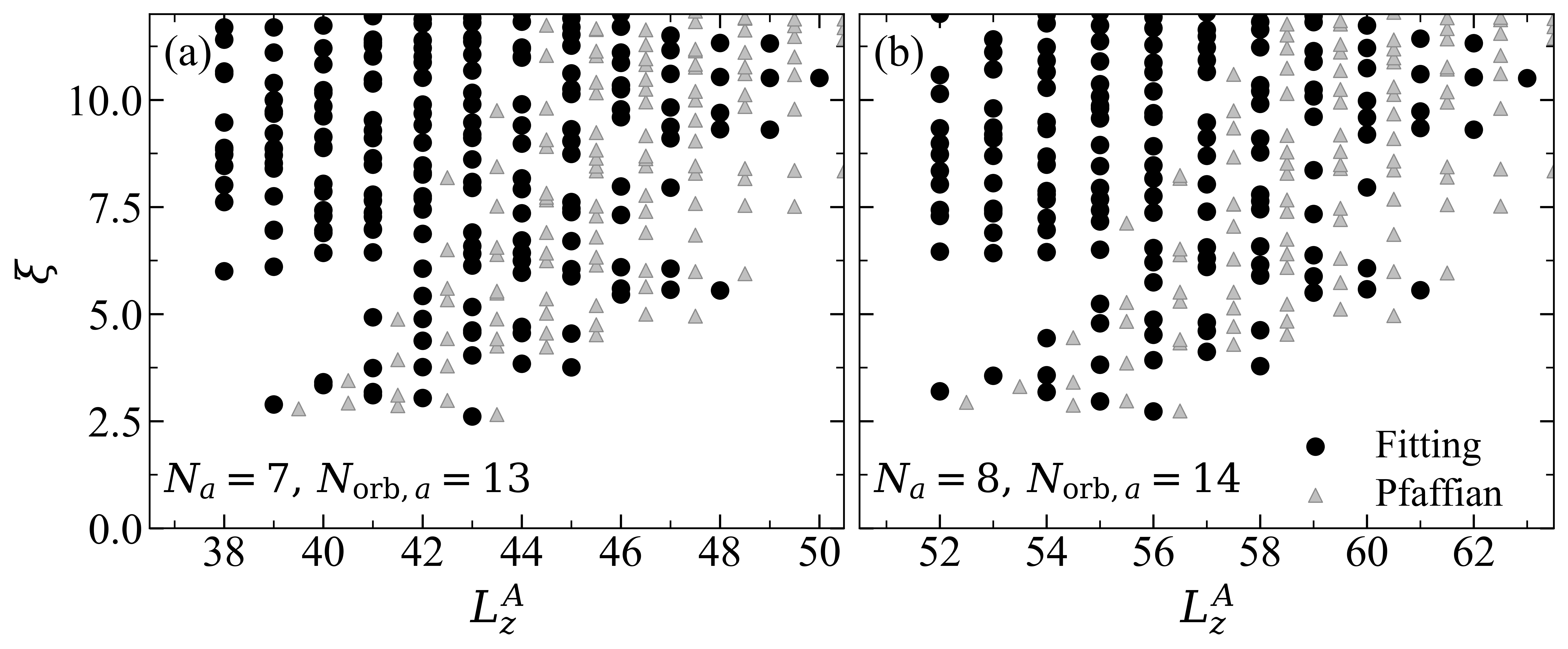}
\par\smallskip
{\small Fig.~\ref{fig:OES14}\par}
\end{figure}

The OES provides a complementary finite-size test of the fitted interaction. As shown in Fig.~\ref{fig:OES14}, the low-lying entanglement levels reproduce the expected Pfaffian counting in each $L_z^A$ sector for both partitions and remain separated from higher nonuniversal levels by a pronounced entanglement gap. For comparison, we also show the OES of the exact MR Pfaffian model wave function, which contains only the universal conformal field theory counting.

These results highlight two aspects of the inverse-designed interaction. First, direct optimization yields an almost exact finite-size realization of the MR Pfaffian state for systems up to $N_e=12$. Second, the extrapolated exponential pseudopotential profile retains the characteristic entanglement structure of the Pfaffian phase at $N_e=14$ without further parameter adjustment. The fitted profile therefore offers a compact parametrization of the interaction, while the truncation test shows that retaining only its leading pseudopotential channels is insufficient to preserve the high overlaps obtained from the full optimized profiles.

Geometric constructions extend pseudopotential Hamiltonians to geometries such as the cylinder and torus, where relative angular momentum is not an exact quantum number~\cite{Lee2015Geometric}. We further assess the fitted interaction on square and rectangular tori for $N_e=8,10$, and $12$ without reoptimization. Because finite-size spectra depend on geometry, transferring the fitted pseudopotential coefficients without reoptimization tests their portability rather than their optimality on the torus. These calculations reveal size- and geometry-dependent transferability and distinguish improvements in MR overlap from those in low-energy spectral isolation (Supplementary Note 6 and Supplementary Figures 6--8). An exploratory direct optimization on the torus further illustrates how spectral constraints can complement a single-sector overlap objective, with the resulting low-energy structure tested under small changes in aspect ratio (Supplementary Figure 9).

\subsubsection*{Real-space interpretation and constrained interactions}

Having tested the fitted interaction across system sizes, we examine its real-space interpretation. In the planar lowest Landau level, the repulsive Gaussian interaction $U(r)=U_0e^{-r^2/(2\sigma^2)}$ generates
\begin{equation}
\label{eq:gaussian_pps}
V_m=U_0\left(\frac{\sigma^2}{\sigma^2+2\ell_B^2}\right)^{m+1}.
\end{equation}
The exponential sequence in Eq.~\eqref{eqfit} is therefore reproduced by $\sigma/\ell_B=\sqrt{2/(e^\alpha-1)}$ and $U_0=Ae^\alpha$, giving $\sigma\simeq3.224\ell_B$ and $U_0\simeq0.66896e^2/(\epsilon\ell_B)$. This interaction is positive, decreases monotonically with separation, and vanishes at infinity. The correspondence applies to the fitted sequence under planar 0LL projection, rather than to an exact inversion of the independently optimized finite-size spherical coefficients. The Fourier transform, projection, and interaction profiles are given in Supplementary Note 4 and Supplementary Figure 4.

Restricting the interaction to a symmetric gate-screening model yields a modest overlap improvement at $N_e=12$, from $0.83575$ to $0.85675$, using planar 1LL pseudopotentials for both the screened and unscreened interactions in spherical exact diagonalization. This illustrates the additional restrictions imposed by this one-parameter family (Supplementary Note 5 and Supplementary Figure 5).

\section*{Conclusion and Outlook}
\label{sec:discussion}

We have developed a differentiable implementation of inverse Hamiltonian design that combines automatic differentiation with exact diagonalization, using the search for two-body interactions approximating the Moore--Read Pfaffian state as a benchmark. By treating the Haldane pseudopotentials as continuous variational parameters and optimizing an overlap-based objective function through differentiable exact diagonalization implemented in JAX, we directly identify interaction profiles whose ground states approach the MR Pfaffian wave function. For systems up to $N_e=12$, the optimized interactions achieve overlaps exceeding $99\%$, demonstrating that appropriately engineered two-body interactions can closely approximate the physics of a non-Abelian FQH state conventionally described by a three-body parent Hamiltonian.

The independently optimized profiles for $N_e=6,8,10$, and $12$ admit a compact common exponential fit within the explored pseudopotential family. Applying this fixed profile to $N_e=14$ retains a Pfaffian overlap of $91.1\%$ without further optimization. Excitation and orbital entanglement spectra provide complementary finite-size diagnostics: the optimized interaction has an isolated $L=0$ ground state and characteristic Pfaffian entanglement counting, while the fixed fitted interaction yields a larger density-corrected neutral gap than the spherical 1LL Coulomb reference over $N_e=8$--$14$ under the stated conventions. Torus calculations further characterize the geometry dependence of the transferred interaction and highlight the complementary information provided by overlaps and low-energy spectra. The smaller higher-$m$ coefficients remain important, as shown by the loss of overlap upon direct truncation. The particle--hole-conjugate anti-Pfaffian calculation provides a numerical consistency check rather than an independent optimization result (Supplementary Note 3 and Supplementary Figures 2 and 3).

The fitted profile also has a simple repulsive Gaussian representation under planar 0LL projection, providing a real-space interpretation of its exponential pseudopotential sequence. The gate-screening test complements this construction by quantifying the overlap improvement attainable within a one-parameter interaction family. Together, these results connect optimization in pseudopotential space with explicitly specified real-space interactions.

Beyond the Moore--Read benchmark, the framework provides a route to searching for approximate two-body parent Hamiltonians for numerically accessible trial states without known parent Hamiltonians. Extending the optimization to larger systems and physically constrained interaction families offers a path toward assessing thermodynamic stability and experimental implementation.

\acknowledgments

We are grateful to Ziang Wang for valuable advice on using the JAX automatic differentiation framework and for guidance on associated coding practices. Yi Yang thanks Chen-Xin Jiang for useful discussions.

\section*{Funding Statement}
This work was supported by the National Natural Science Foundation of China (Grant Nos. 12474140 and 12547101) and the Fundamental Research Funds for the Central Universities (Grant No. 2025CDJ-IAISYB-029).

\section*{Data Availability}
The numerical source data underlying the figures and tables in this study are publicly available on Zenodo~\cite{Yang2026SourceData} at \url{https://doi.org/10.5281/zenodo.23838158}.

\section*{Code Availability}
The custom code used in this study is available from the corresponding authors upon request. The torus transferability calculations and three-body MR reference states were obtained using DiagHam~\cite{DiagHam}, with a custom interaction kernel for the fitted pseudopotentials. Direct two-body optimization on the torus used our own magnetic-translation-resolved implementation.

\section*{Competing Interests}
The authors declare no competing interests.

\section*{Author Contributions}
Y.Y. implemented the numerical framework, performed all simulations, analyzed the data, and wrote the manuscript. X.W. and Z.-X.H. conceived the project, supervised the research, and revised the manuscript. All authors discussed the results and commented on the manuscript.

\bibliography{biblio_fqhe.bib}

\clearpage
\onecolumngrid
\section*{Figure legends}

\begin{figure}[H]
\caption{\label{model}Differentiable inverse Hamiltonian design for the Moore--Read state. \textbf{a}, Schematic of electrons on a sphere. \textbf{b}, Construction of the occupation-number basis in the $L_z=0$ sector, with $N_e$ electrons and $N_{\mathrm{orb}}=2N_e-2$ orbitals; 1 and 0 denote occupied and empty orbitals, respectively. \textbf{c}, Preparation of the Moore--Read Pfaffian target state using a Jack polynomial or a three-body parent Hamiltonian. \textbf{d,e}, Construction and exact diagonalization of the two-body Hamiltonian, where $V_m$ is the Haldane pseudopotential and $P_m=\sum_{i<j}\hat{P}_{ij}^{(m)}$ sums the pair projectors in relative-angular-momentum channel $m$. \textbf{f,g}, Evaluation of the overlap-based loss for normalized states and its gradient with respect to the pseudopotentials using automatic differentiation in JAX. \textbf{h}, Sequential least-squares programming (SLSQP) updates the pseudopotentials subject to nonnegativity and monotonicity, with $V_1$ fixed to the spherical lowest-Landau-level Coulomb value. The optimization loop is repeated until convergence, after which the optimized interaction is assessed through excitation and entanglement spectra.}
\end{figure}

\begin{figure}[H]
\caption{Optimized Haldane pseudopotentials and Coulomb reference interactions. The system contains $N_e=12$ electrons in 22 orbitals at the Moore--Read shift. The pseudopotential $V_m$ specifies the interaction strength in pair relative-angular-momentum channel $m$. Black open circles connected by a solid line denote the optimized interaction; blue open squares connected by a dashed line and red open triangles connected by a dash-dotted line denote the Coulomb interactions projected into the lowest (0LL) and first excited (1LL) Landau levels, respectively. The optimized interaction exhibits a shorter-ranged pseudopotential profile and yields a $99.49\%$ overlap with the Moore--Read Pfaffian state. The 0LL and 1LL Coulomb values of $V_1$ differ by approximately $0.75\%$ and therefore appear nearly coincident at the scale shown; neither Coulomb reference has been rescaled to a common $V_1$.}
\label{fig:pps12}
\end{figure}

\begin{figure}[H]
\caption{Neutral excitation spectra of the optimized and Coulomb interactions. The system contains $N_e=12$ electrons in $N_{\mathrm{orb}}=22$ orbitals on the sphere. Black open circles represent the optimized interaction, while black crosses represent the Coulomb interaction projected into the first excited Landau level (1LL). The horizontal coordinate is $k=L/R$, where $L$ is the total angular momentum and $R=\sqrt{Q}\ell_B$ is the sphere radius for the respective interaction; $k$ is expressed in units of $\ell_B^{-1}$. The different monopole strengths therefore produce slightly different horizontal positions at the same $L$. The plotted excitation energies include the finite-size density correction $\Delta E_{\mathrm{corr}}=\sqrt{2Q\nu/N_e}(E-E_0)$, where $E$ and $E_0$ are the eigenstate and ground-state energies of the respective interactions, $\nu=1/2$ is the filling factor, and $Q$ is the monopole strength. We use $Q=10.5$ for the optimized interaction represented in the lowest Landau level (0LL) and $Q=9.5$ for the 1LL Coulomb reference, following $N_{\mathrm{orb}}=2(Q+n)+1$ for Landau level index $n$. Energies are expressed in units of $e^2/(\epsilon\ell_B)$, where $e$ is the elementary charge, $\epsilon$ is the dielectric constant, and $\ell_B$ is the magnetic length. Under these conventions, the optimized interaction exhibits a larger finite-size neutral gap than the 1LL Coulomb reference.}
\label{fig:energy}
\end{figure}

\begin{figure}[H]
\caption{Orbital entanglement spectra of the optimized and Moore--Read states. The system contains $N_e=12$ electrons in 22 orbitals on the sphere at the Moore--Read (MR) shift. Subsystem $A$ consists of $N_{\mathrm{orb},a}$ contiguous orbitals, and $N_a$ denotes the number of electrons in that subsystem. The partitions are \textbf{a}, $N_a=6$, $N_{\mathrm{orb},a}=11$; and \textbf{b}, $N_a=7$, $N_{\mathrm{orb},a}=12$. Black circles represent the optimized ground state, while grey triangles represent the exact MR Pfaffian state. The horizontal coordinate $L_z^A$ is the angular-momentum component of subsystem $A$, and $\xi$ denotes the entanglement energy. For clarity, the Pfaffian data are shifted horizontally by $0.5$. In both partitions, the optimized state reproduces the Pfaffian counting and low-lying dispersion, with an entanglement gap separating these levels from higher nonuniversal levels.}
\label{fig:OES}
\end{figure}

\begin{figure}[H]
\caption{Orbital entanglement spectra of the optimized and Coulomb ground states. The system contains $N_e=12$ electrons in 22 orbitals on the sphere at the Moore--Read shift. Black circles denote the optimized state, and grey triangles denote the Coulomb ground state in the first excited Landau level (1LL). Subsystem $A$ contains $N_a$ electrons in $N_{\mathrm{orb},a}$ contiguous orbitals: \textbf{a}, $N_a=6$, $N_{\mathrm{orb},a}=11$; \textbf{b}, $N_a=7$, $N_{\mathrm{orb},a}=12$. Here, $L_z^A$ is the subsystem angular-momentum component and $\xi$ is the entanglement energy. The Coulomb data are shifted horizontally by $0.5$ for clarity.}
\label{fig:comparison1LLOES}
\end{figure}

\begin{figure}[H]
\caption{Finite-size dependence of the optimized pseudopotentials. Independently optimized Haldane pseudopotentials $V_m$ are shown for systems containing $N_e=6,8,10$, and $12$ electrons on the sphere at the Moore--Read shift, where $m$ is the pair relative angular momentum. Orange open squares, green stars, blue crosses, and pink plus signs denote $N_e=6,8,10$, and $12$, respectively. The gold solid curve shows the common exponential fit $V_m\simeq0.561e^{-0.176m}$. Monotonicity is imposed during optimization, whereas the exponential form is fitted afterwards.}
\label{fig:fitpps}
\end{figure}

\begin{figure}[H]
\caption{Finite-size dependence of the neutral gap and Pfaffian overlap. The electron numbers are $N_e=8,10,12$, and $14$. Solid lines show the density-corrected neutral gap $\Delta E_{\mathrm{corr}}$ on the left axis, while dashed lines show the overlap magnitude with the Moore--Read (MR) Pfaffian state on the right axis. Black open circles denote the fixed fitted interaction $V_m=0.561e^{-0.176m}$, where $m$ is the pair relative angular momentum; black crosses denote the spherical Coulomb interaction in the first excited Landau level (1LL). The fitted coefficients are used without further optimization or size-dependent rescaling. Gaps include the finite-size density correction described in the text and are expressed in units of $e^2/(\epsilon\ell_B)$, where $e$ is the elementary charge, $\epsilon$ is the dielectric constant, and $\ell_B$ is the magnetic length. Connecting lines are guides to the eye, not extrapolation fits.}
\label{fig:gapoverlap}
\end{figure}

\begin{figure}[H]
\caption{Orbital entanglement spectra for a larger system. The system contains $N_e=14$ electrons in 26 orbitals on the sphere at the Moore--Read shift. Black circles denote the ground state obtained from the fixed fitted pseudopotentials $V_m=0.561e^{-0.176m}$ without further optimization, where $m$ is the pair relative angular momentum; grey triangles denote the exact Moore--Read Pfaffian state. Subsystem $A$ contains $N_a$ electrons in $N_{\mathrm{orb},a}$ contiguous orbitals: \textbf{a}, $N_a=7$, $N_{\mathrm{orb},a}=13$; \textbf{b}, $N_a=8$, $N_{\mathrm{orb},a}=14$. Here, $L_z^A$ is the subsystem angular-momentum component and $\xi$ is the entanglement energy. The Pfaffian data are shifted horizontally by $0.5$ for clarity.}
\label{fig:OES14}
\end{figure}

\end{document}


\title{Supplementary Information for ``Differentiable Inverse Design of Two-Body Interactions for the Moore--Read State''}
\author{Yi Yang, Xin Wan, and Zi-Xiang Hu}
\maketitle

\section*{Supplementary Note 1. Topological entanglement entropy}

\begin{figure}[H]
\centering
\includegraphics[width=0.65\textwidth]{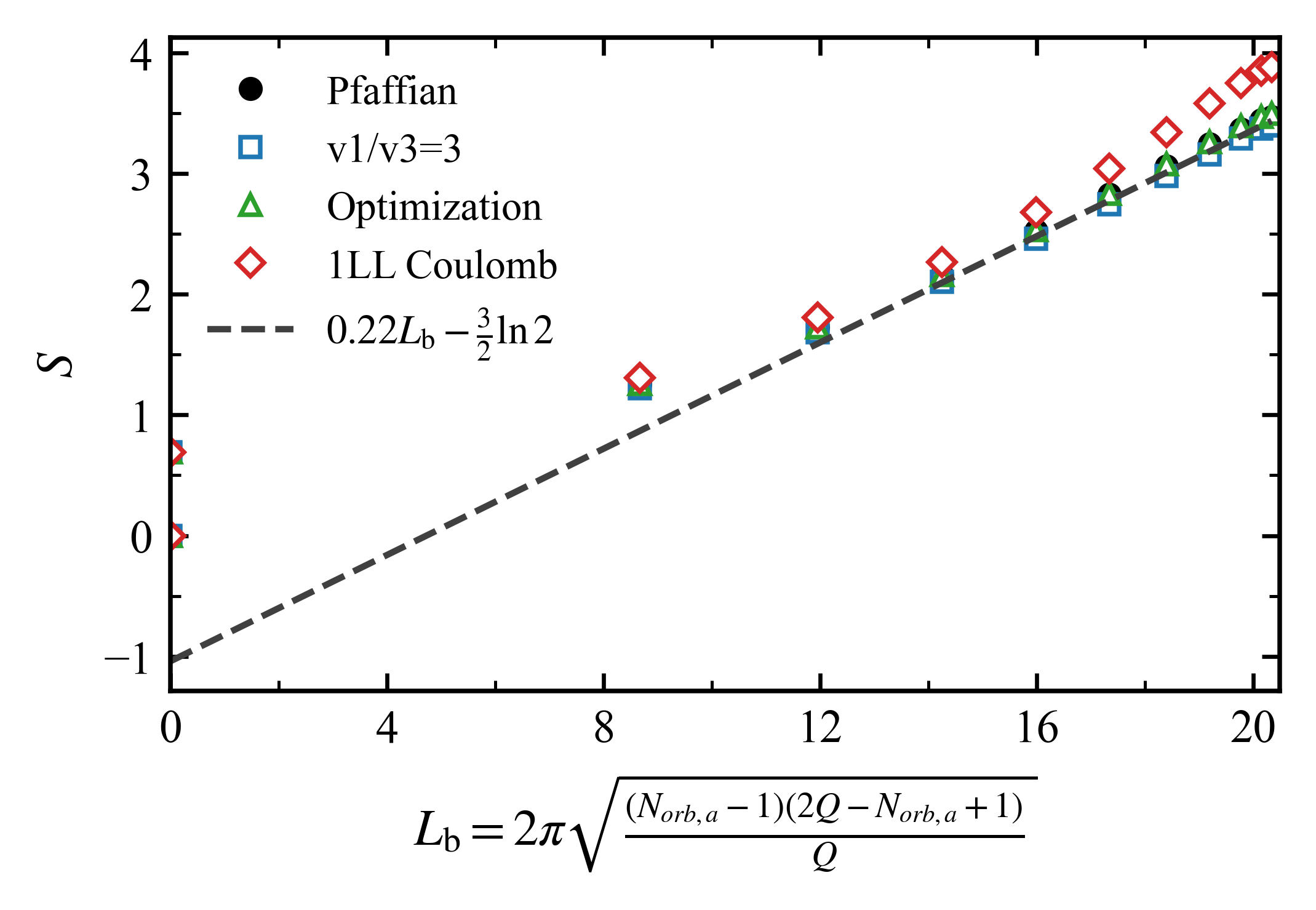}
\caption{Orbital entanglement entropy of the optimized and reference states. The system contains $N_e=12$ electrons in $N_{\mathrm{orb}}=22$ orbitals on the sphere. The entanglement entropy $S$ is plotted against the orbital-cut boundary length $L_{\mathrm{b}}$. Here, $N_{\mathrm{orb},a}$ is the number of contiguous orbitals in subsystem $A$, and $Q$ is the monopole strength used in the boundary-length formula. Green triangles denote the optimized ground state, black circles denote the exact Moore--Read Pfaffian state, blue squares denote the particle--hole-symmetric reference state from Ref.~\onlinecite{V1V33}, and red diamonds denote the Coulomb ground state in the first excited Landau level (1LL). The grey dashed line shows the reference area-law form $S=0.22L_{\mathrm{b}}-\ln\sqrt{8}$.}
\label{fig:entropy}
\end{figure}

The topological entanglement entropy (TEE) provides a characterization of long-range entanglement associated with topological order. We compute the orbital-cut entanglement entropy on the sphere for $N_e=12$ and $N_{\mathrm{orb}}=22$ using the standard bipartition protocol. As shown in Supplementary Figure~\ref{fig:entropy}, the optimized pseudopotential state exhibits an entanglement entropy nearly identical to that of the exact MR Pfaffian state over the accessible range of subsystem sizes. This agreement provides an additional finite-size comparison of the optimized and Pfaffian states.

For comparison, we also consider the ground state of the Coulomb interaction projected into the 1LL and the particle-hole-symmetric reference state introduced in Ref.~\onlinecite{V1V33}. The dashed line shows the reference area-law form $S=0.22L_{\mathrm{b}}-\ln\sqrt{8}$, with the Pfaffian topological contribution $\gamma=\ln\sqrt{8}$. The optimized and exact Pfaffian states exhibit closely matching entropy profiles. In contrast, the particle-hole-symmetric reference state shows prominent finite-size deviations at some subsystem sizes, while the 1LL Coulomb ground state displays systematically larger entanglement entropy values, lying above the reference area-law curve.

\clearpage
\section*{Supplementary Note 2. Pfaffian overlaps from the fitted pseudopotentials}
\label{sec:appendix}

We further examine the overlap $\mathcal{O}=|\langle\Psi_{\rm Pfaffian}|\Psi_0\rangle|$ between the MR Pfaffian state and ground states generated by different interaction models. In particular, we evaluate the interaction defined by the fixed fitted pseudopotential profile $V_m=0.561e^{-0.176m}$ in units of $e^2/(\epsilon\ell_B)$ and compare it with the Coulomb interaction projected into the 0LL and 1LL.

The results are summarized in Supplementary Table~\ref{tab:overlap}. The fitted pseudopotential profile maintains large Pfaffian overlaps for the system sizes listed in Supplementary Table~\ref{tab:overlap}, despite the absence of additional size-dependent optimization. The 1LL Coulomb interaction also exhibits sizable overlaps, whereas the LLL Coulomb interaction shows substantially weaker overlap, consistent with the compressible composite-fermion liquid behavior expected at half filling in the LLL.

\begin{table}[H]
\centering
\caption{Pfaffian overlaps for different interaction models. Entries give the overlap magnitude between the normalized Moore--Read Pfaffian state and the corresponding ground state for $N_e$ electrons. ``Optimization'' denotes independent optimization at each system size, while ``Fitting'' denotes the fixed exponential pseudopotential profile without further optimization. The 0LL and 1LL Coulomb columns refer to interactions projected into the lowest and first excited Landau levels, respectively. A dash indicates that direct optimization was not performed.}
\label{tab:overlap}
\begingroup
\large
\renewcommand{\arraystretch}{1.35}
\setlength{\tabcolsep}{12pt}
\begin{tabular}{ccccc}
\toprule
$N_e$ & Optimization & Fitting & 0LL Coulomb & 1LL Coulomb \\
\midrule
10 & 0.9998 & 0.95549 & 0 & 0.83764 \\
12 & 0.9949 & 0.92896 & 0.65519 & 0.81938 \\
14 & -- & 0.91089 & 0.72226 & 0.69345 \\
\bottomrule
\end{tabular}
\endgroup
\end{table}

\clearpage
\section*{Supplementary Note 3. Particle--hole symmetry and the anti-Pfaffian state}
\label{sec:ph_consistency}

The Pfaffian and anti-Pfaffian states are related by particle--hole conjugation within a single Landau level. At $N_\phi=21$, there are $N_{\mathrm{orb}}=22$ orbitals, and the particle--hole conjugate of the $N_e=12$ Pfaffian state contains $N_e=10$ electrons. We construct this anti-Pfaffian target by replacing each orbital occupation $n_j$ with $1-n_j$ in the Pfaffian basis configurations. For the real wave-function coefficients and fixed angular-momentum sector considered here, this transformation leaves the coefficients unchanged up to an overall phase.

For a rotationally invariant two-body interaction projected into a single Landau level, particle--hole conjugation preserves the two-body part of the Hamiltonian and generates only a uniform one-body term and an additive constant. Within each fixed-particle-number sector, these terms shift the energy without changing the eigenstates. Consequently, for the same pseudopotential sequence $\mathbf{V}=\{V_m\}$, the nondegenerate ground states in the two conjugate sectors satisfy
\begin{equation}
|\langle\Psi_{\mathrm{aPf}}^{(10)}|\Psi_0^{(10)}(\mathbf{V})\rangle|=|\langle\Psi_{\mathrm{Pf}}^{(12)}|\Psi_0^{(12)}(\mathbf{V})\rangle|.
\label{eq:ph_overlap}
\end{equation}
Here, the superscripts indicate the electron number, and both sectors have $N_\phi=21$. The Pfaffian and anti-Pfaffian overlap objectives are therefore equivalent when the same parameter domain and constraints are used.

\begin{figure}[H]
\centering
\includegraphics[width=0.65\textwidth]{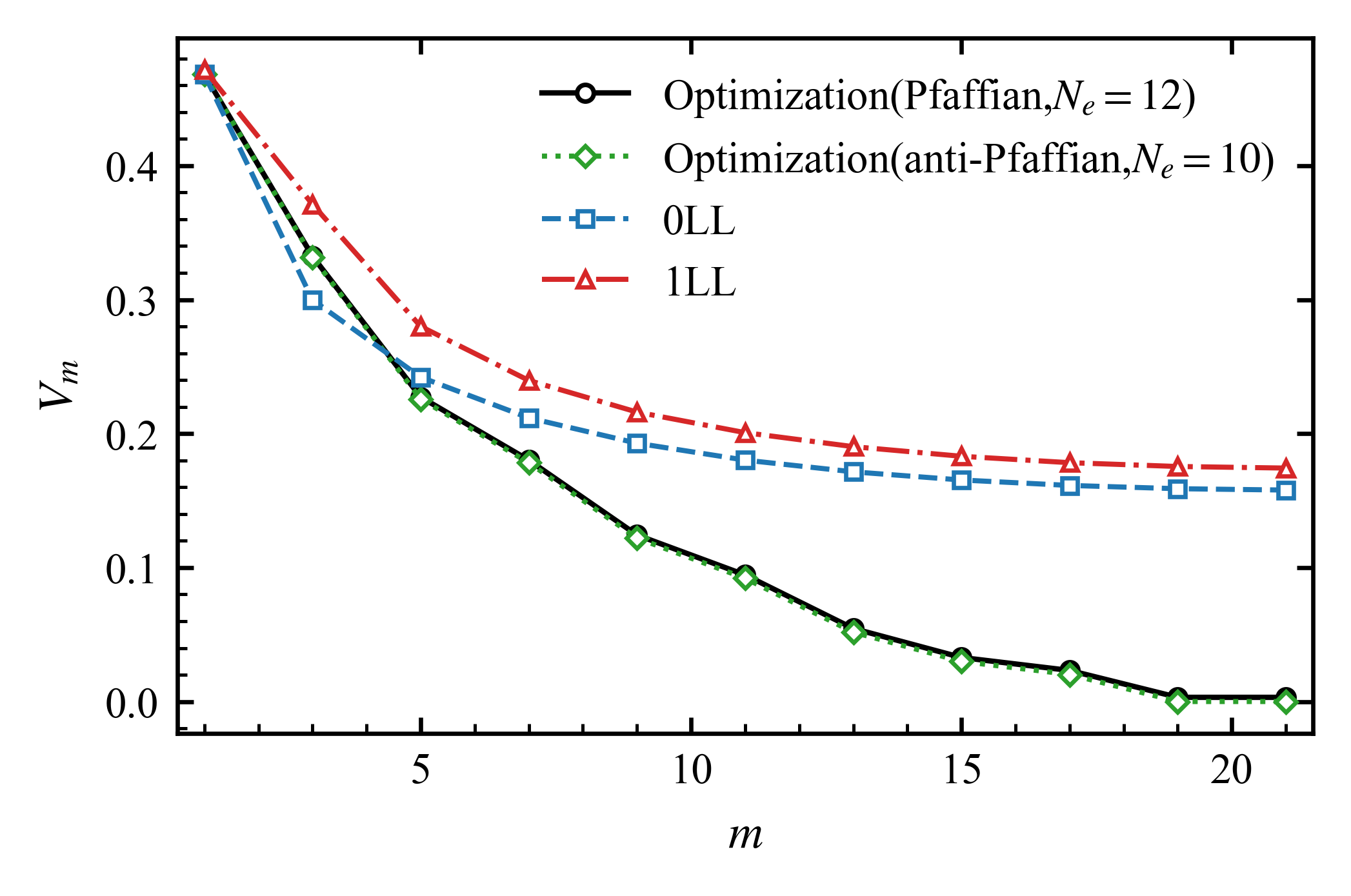}
\caption{Pseudopotentials from Pfaffian and anti-Pfaffian optimizations. Black open circles connected by a solid line denote the interaction optimized for the Moore--Read Pfaffian state with $N_e=12$ electrons; green open diamonds connected by a dotted line denote the interaction optimized for its particle--hole-conjugate anti-Pfaffian state with $N_e=10$ electrons. Both systems contain 22 orbitals. The Haldane pseudopotential $V_m$ specifies the interaction strength in pair relative-angular-momentum channel $m$. Blue open squares connected by a dashed line and red open triangles connected by a dash-dotted line denote the Coulomb references in the lowest (0LL) and first excited (1LL) Landau levels, respectively. Both optimizations yield an overlap of $99.49\%$ at the reported precision, consistent with the particle--hole relation in Eq.~(\ref{eq:ph_overlap}).}
\label{fig:ppsapf}
\end{figure}

As a numerical consistency check, we performed a separate optimization with the anti-Pfaffian target. Supplementary Figure~\ref{fig:ppsapf} compares the resulting pseudopotentials with those obtained from the Pfaffian optimization. The anti-Pfaffian overlap is $99.49\%$, equal to the Pfaffian value at the reported precision. This agreement is expected from particle--hole symmetry and does not constitute an independent optimization result. The similar profiles obtained in the two runs are consistent with their equivalent objectives, although this equivalence does not imply that the optimal pseudopotential sequence is unique.

\begin{figure}[H]
\centering
\includegraphics[width=0.65\textwidth]{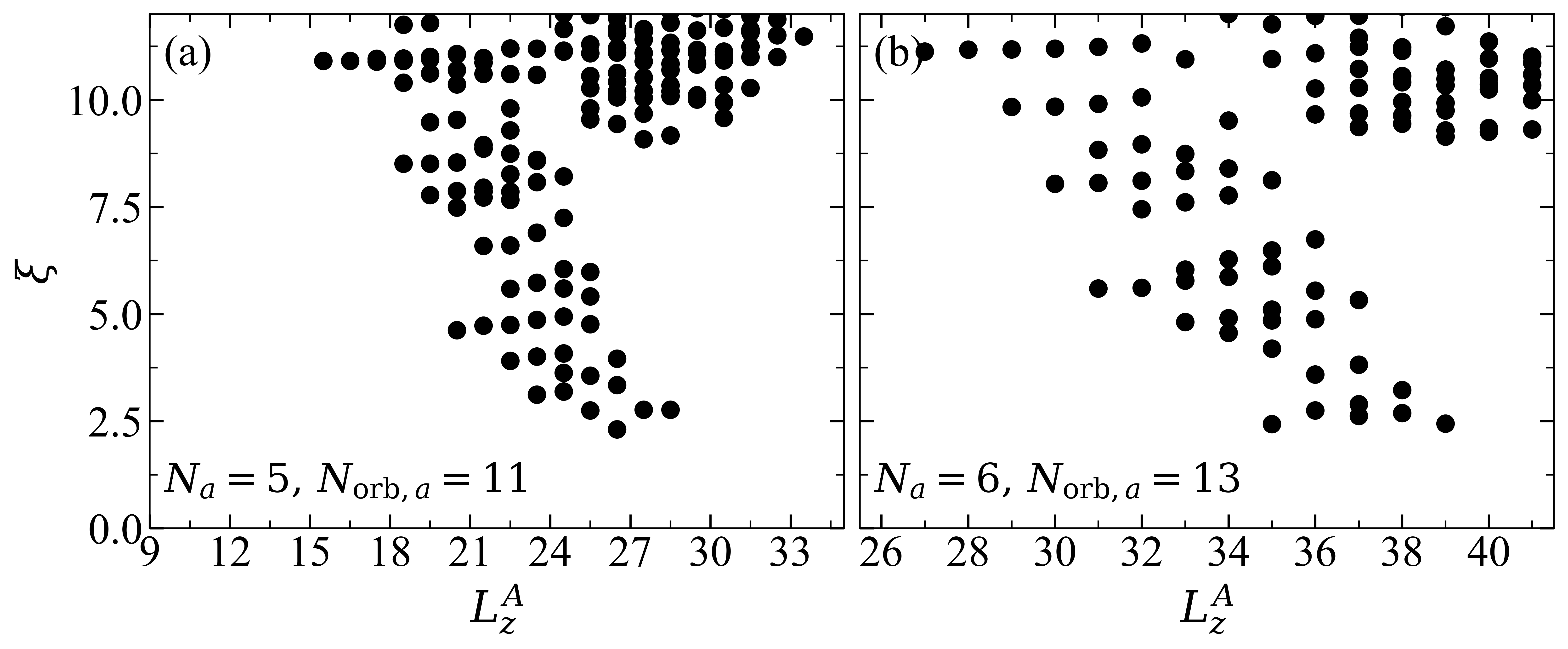}
\caption{Orbital entanglement spectra of the ground state optimized toward the anti-Pfaffian target. Black circles show the entanglement levels for a system containing $N_e=10$ electrons in 22 orbitals on the sphere. Subsystem $A$ contains $N_a$ electrons in $N_{\mathrm{orb},a}$ contiguous orbitals: \textbf{a}, $N_a=5$, $N_{\mathrm{orb},a}=11$; \textbf{b}, $N_a=6$, $N_{\mathrm{orb},a}=13$. Here, $L_z^A$ is the subsystem angular-momentum component and $\xi$ is the entanglement energy. In both partitions, the low-lying levels exhibit anti-Pfaffian counting and are separated from higher nonuniversal levels by an entanglement gap.}
\label{fig:apfOES}
\end{figure}

Supplementary Figure~\ref{fig:apfOES} shows the OES of the optimized $N_e=10$ state for the orbital partitions $(N_a,N_{\mathrm{orb},a})=(5,11)$ and $(6,13)$. The low-lying levels exhibit the anti-Pfaffian counting and are separated from higher nonuniversal levels by an entanglement gap. These spectra provide a complementary characterization of the state in the particle--hole-conjugate sector.

\clearpage
\section*{Supplementary Note 4. Gaussian representation of the fitted pseudopotentials}

\begin{figure}[H]
\centering
\includegraphics[width=\textwidth]{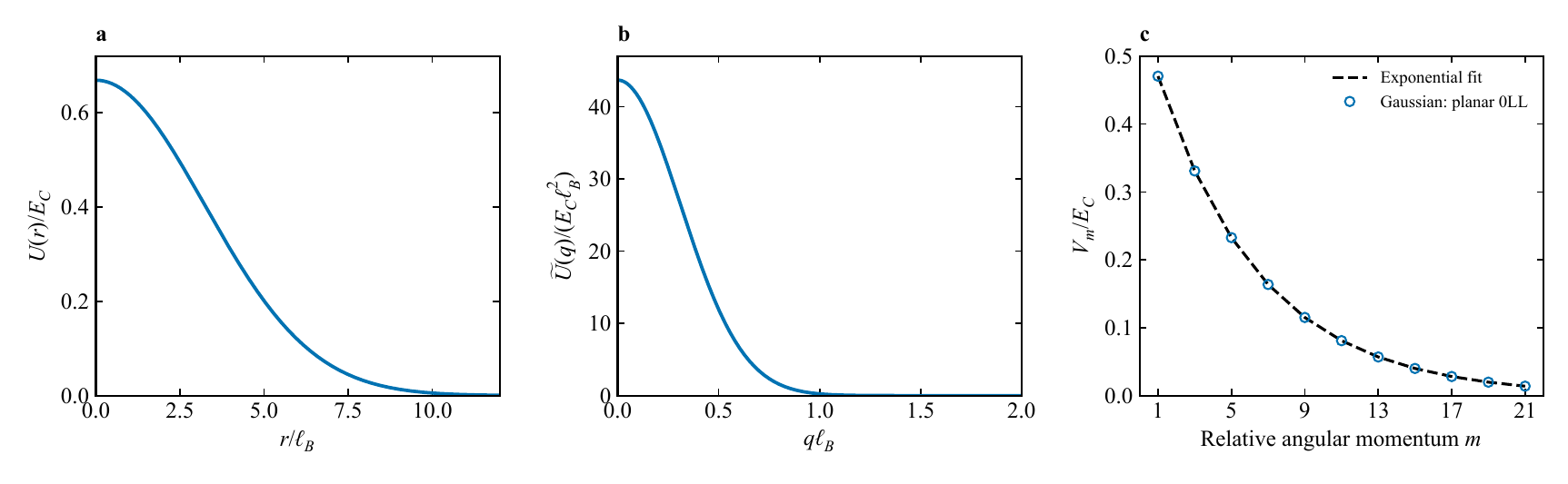}
\caption{Repulsive Gaussian representation of the fitted exponential pseudopotentials. \textbf{a}, Real-space interaction $U(r)=U_0e^{-r^2/(2\sigma^2)}$. \textbf{b}, Its two-dimensional Fourier transform $\widetilde{U}(q)$. \textbf{c}, Odd pseudopotentials obtained by numerical planar lowest-Landau-level (0LL) projection of the Gaussian interaction (blue open circles), compared with the exponential fit $V_m/E_C=0.561e^{-0.176m}$ (black dashed line). Here, $r$ is the particle separation, $q$ is the wave-number magnitude, $m$ is the relative angular momentum, $\ell_B$ is the magnetic length, and $E_C=e^2/(\epsilon\ell_B)$, where $\epsilon$ is the dielectric constant. The parameters are $U_0/E_C=0.561e^{0.176}$ and $\sigma/\ell_B=\sqrt{2/(e^{0.176}-1)}$. The two pseudopotential sequences agree within numerical precision.}
\label{fig:gaussian_representation}
\end{figure}

The fitted sequence $V_m=Ae^{-\alpha m}$ admits an exact representation under planar lowest-Landau-level (0LL) projection. With the Fourier convention $\widetilde{U}(q)=\int d^2r\,e^{-i\mathbf{q}\cdot\mathbf{r}}U(r)$, the Gaussian interaction $U(r)=U_0e^{-r^2/(2\sigma^2)}$ has transform $\widetilde{U}(q)=2\pi U_0\sigma^2e^{-q^2\sigma^2/2}$. Its planar 0LL pseudopotentials are
\begin{equation}
V_m=\int_0^\infty\frac{q\,dq}{2\pi}\widetilde{U}(q)L_m(q^2\ell_B^2)e^{-q^2\ell_B^2}=U_0\left(\frac{\sigma^2}{\sigma^2+2\ell_B^2}\right)^{m+1},
\end{equation}
where $L_m$ is a Laguerre polynomial. Choosing $\sigma/\ell_B=\sqrt{2/(e^\alpha-1)}$ and $U_0=Ae^\alpha$ reproduces the fitted sequence. For $A=0.561E_C$ and $\alpha=0.176$, with $E_C=e^2/(\epsilon\ell_B)$, these parameters are $\sigma\simeq3.224\ell_B$ and $U_0\simeq0.66896E_C$. Supplementary Figure~\ref{fig:gaussian_representation} shows the real-space interaction, its Fourier transform, and the agreement between the numerically projected odd pseudopotentials and the exponential fit. The Gaussian also specifies even pseudopotentials, which do not contribute for fully spin-polarized fermions.

This interaction is everywhere positive, decreases monotonically with separation, and vanishes at infinity. The exact correspondence concerns the fitted sequence under planar 0LL projection, not an exact inversion of the independently optimized finite-size spherical coefficients. We use planar projection to provide a size-independent real-space interpretation of the common fitted sequence. For the same interaction at fixed $\sigma/\ell_B$, spherical pseudopotentials generally exhibit finite-size corrections and approach their planar values at fixed relative angular momentum $m$ in the thermodynamic limit.

\clearpage
\section*{Supplementary Note 5. Gate-screened interaction}

We test the symmetric gate-screening model $V_d(q)=2\pi e^2\tanh(qd)/(\epsilon q)$, where $q$ is the in-plane wave-number magnitude and $d$ is the distance from the electron layer to either of two symmetrically placed gates. Using $k=q\ell_B$ and $\delta=d/\ell_B$, its planar first-excited-Landau-level (1LL) pseudopotentials are
\begin{equation}
\frac{V_m^{(1)}(d)}{E_C}=\int_0^\infty dk\,\tanh(k\delta)\left(1-\frac{k^2}{2}\right)^2L_m(k^2)e^{-k^2},
\end{equation}
where $E_C=e^2/(\epsilon\ell_B)$ and $L_m$ is a Laguerre polynomial. We use the odd channels $m=1,3,\ldots,21$ in spherical exact diagonalization for $N_e=12$ electrons in 22 orbitals. Only $d/\ell_B$ is varied; all coefficients, including $V_1$, follow from the screened interaction without additional rescaling. The unscreened reference is obtained by taking $d/\ell_B\to\infty$ in the same planar formula. This comparison therefore uses planar 1LL pseudopotentials for both interactions, rather than the finite-sphere Coulomb coefficients used elsewhere in the manuscript.

\begin{figure}[H]
\centering
\includegraphics[width=0.72\textwidth]{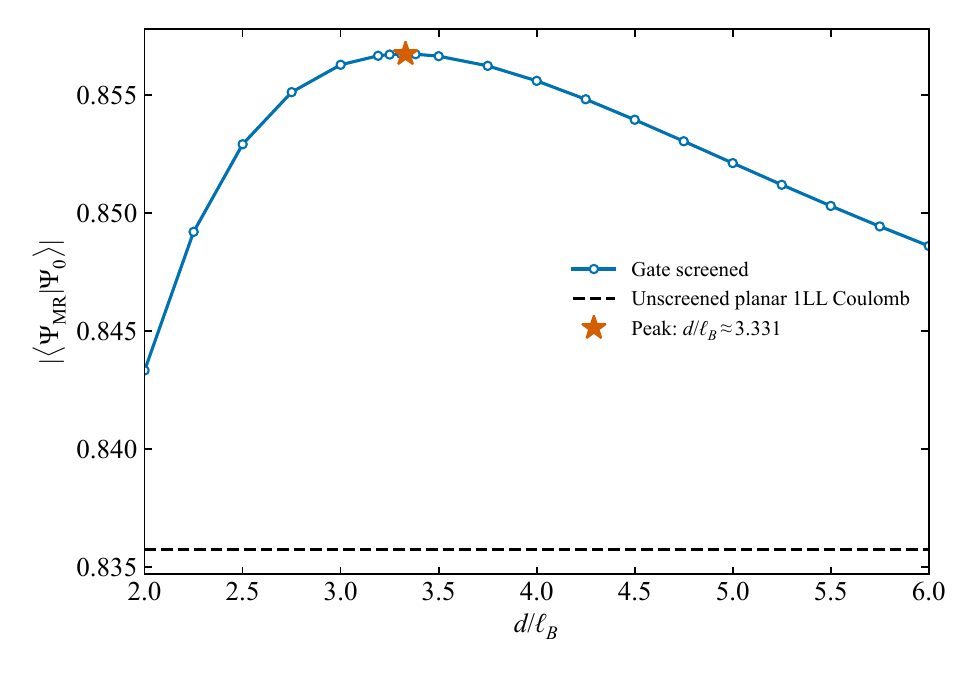}
\caption{Pfaffian overlap for a gate-screened interaction. Ground-state overlap magnitude $|\langle\Psi_{\mathrm{MR}}|\Psi_0\rangle|$ for $N_e=12$ electrons in 22 orbitals, calculated on the sphere using planar first-excited-Landau-level (1LL) pseudopotentials. Here, $\Psi_{\mathrm{MR}}$ is the normalized Moore--Read Pfaffian state and $\Psi_0$ is the normalized ground state. Blue open circles show the screened results, the black dashed line indicates the unscreened planar 1LL Coulomb reference, and the orange star marks the peak located near $d/\ell_B=3.33$, where $d$ is the distance to each symmetric gate and $\ell_B$ is the magnetic length. All displayed ground states have total angular momentum $L=0$. Connecting lines are guides to the eye.}
\label{fig:gate_screening}
\end{figure}

An initial scan at $d/\ell_B=0.25,0.5,1,2,5$, and $10$ was followed by a finer scan over $2\le d/\ell_B\le6$ and refinement near the observed peak. The overlap magnitude reaches approximately $0.85675$ near $d/\ell_B=3.33$, compared with $0.83575$ in the unscreened limit (Supplementary Figure~\ref{fig:gate_screening}). All ground states in the displayed interval have total angular momentum $L=0$. At the stronger-screening points $d/\ell_B=0.25$ and $0.5$, the ground state instead has $L=6$ and is orthogonal to the $L=0$ Pfaffian target. The improvement of approximately $2.10$ percentage points shows that this one-parameter interaction family retains only a modest overlap enhancement and does not reproduce the $99.49\%$ overlap obtained by independently optimizing the pseudopotentials. The calculation quantifies the improvement attainable within this specified screening model.

\clearpage
\section*{Supplementary Note 6. Transferability to the torus}

We examine the transferability of the fixed fitted interaction to the torus at filling $\nu=1/2$, with $N_\phi=2N_e$ orbitals. We consider $N_e=8,10$, and $12$ on square and rectangular tori with aspect ratios $L_x/L_y=1$ and $1.2$, respectively, and area $L_xL_y=2\pi N_\phi\ell_B^2$. The interaction parameters are transferred without reoptimization or size-dependent rescaling. We compare the fitted interaction with the Coulomb interaction projected into the first excited Landau level (1LL) and use the three-body Moore--Read (MR) parent Hamiltonian to define the reference states. The transferability calculations and the three-body MR reference states were obtained using DiagHam~\cite{DiagHam}; the direct two-body optimization described below uses our own magnetic-translation-resolved implementation.

We transfer the fitted pseudopotential profile $V_m=0.561E_Ce^{-0.176m}$ obtained from the spherical optimization to the torus without reoptimization, where $E_C=e^2/(\epsilon\ell_B)$. The interaction matrix elements are evaluated using the analytically summed Fourier kernel corresponding to the full odd-$m$ sequence. The 1LL Coulomb Hamiltonian is constructed directly from the Fourier-space Coulomb interaction with the 1LL form factor, without truncating a pseudopotential expansion. The $q=0$ component is omitted, corresponding to the conventional treatment with a uniform neutralizing background; this does not affect the excitation energies or eigenstates at fixed particle number and area. All spectra are measured relative to the ground-state energy of the respective Hamiltonian. The two-body energies are expressed in units of $E_C$, without a spherical finite-size density correction or rescaling to a common $V_1$. The three-body MR parent spectra are shown in arbitrary energy units using the normalization of the numerical implementation. Their overall energy scale is independent of $E_C$ and is not used for quantitative gap comparisons with the two-body interactions.

We resolve the spectra into magnetic-translation sectors labeled by $K_x=0,\ldots,N_e-1$ and $K_y=0,\ldots,N_\phi-1$, retaining the center-of-mass partners. The three-body parent Hamiltonian has six zero-energy states in the systems studied. For each of their momentum sectors, we calculate the overlap magnitude $|\langle\Psi_{\mathrm{MR}}^{(K_x,K_y)}|\Psi_0^{(K_x,K_y)}\rangle|$, where both states are normalized and $\Psi_0^{(K_x,K_y)}$ is the lowest-energy two-body eigenstate in the same sector. These sector-resolved overlaps do not assume that the six selected states form the global lowest-energy manifold.

\begin{figure}[H]
\centering
\includegraphics[width=\textwidth]{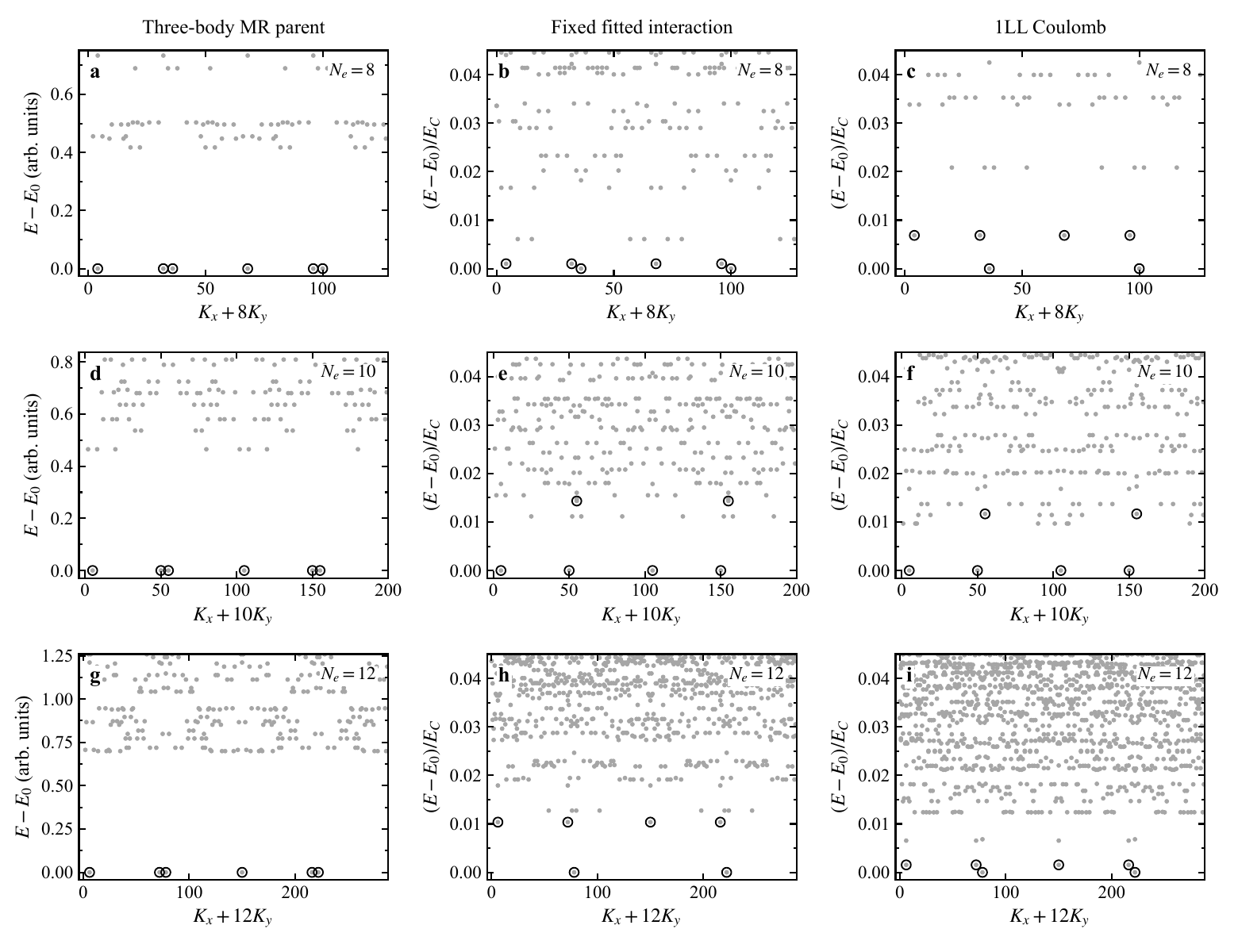}
\caption{Low-energy spectra on the square torus. Rows correspond to $N_e=8$ (\textbf{a--c}), $10$ (\textbf{d--f}), and $12$ (\textbf{g--i}), with $N_\phi=2N_e$ and $L_x/L_y=1$. Columns show the three-body Moore--Read (MR) parent Hamiltonian, the fixed fitted interaction, and the first-excited-Landau-level (1LL) Coulomb interaction, respectively. Grey points show up to eight lowest levels per momentum sector within the displayed energy window. Black open circles mark the lowest state in each of the six momentum sectors containing a zero-energy MR parent state; for the two-body interactions, this marking does not imply an isolated or degenerate six-state manifold. The horizontal coordinate $K_x+N_eK_y$ enumerates the magnetic-translation sectors. Energies are measured relative to each Hamiltonian's ground-state energy $E_0$. The two-body energies are expressed in units of $E_C=e^2/(\epsilon\ell_B)$. The three-body parent spectra are shown in arbitrary energy units, with an independent overall normalization; their gap magnitudes are not compared with those of the two-body interactions.}
\label{fig:torus_square}
\end{figure}

\begin{figure}[H]
\centering
\includegraphics[width=\textwidth]{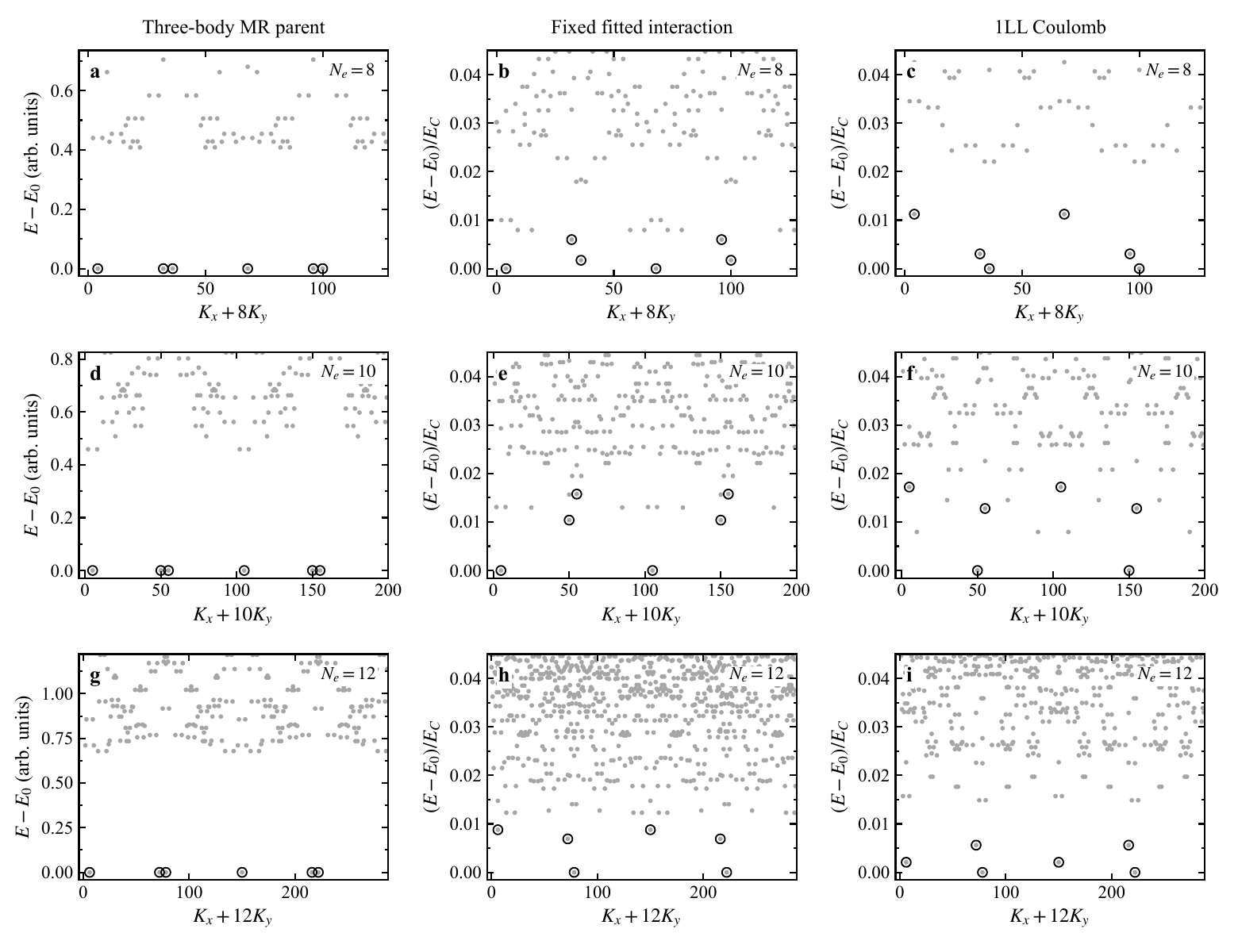}
\caption{Low-energy spectra on the rectangular torus. The aspect ratio is $L_x/L_y=1.2$. Rows correspond to $N_e=8$ (\textbf{a--c}), $10$ (\textbf{d--f}), and $12$ (\textbf{g--i}), with $N_\phi=2N_e$. Columns show the three-body Moore--Read parent Hamiltonian, the fixed fitted interaction, and the 1LL Coulomb interaction, respectively. Symbols, momentum labels, and energy conventions are the same as in Supplementary Figure~\ref{fig:torus_square}.}
\label{fig:torus_rectangular}
\end{figure}

\begin{figure}[H]
\centering
\includegraphics[width=\textwidth]{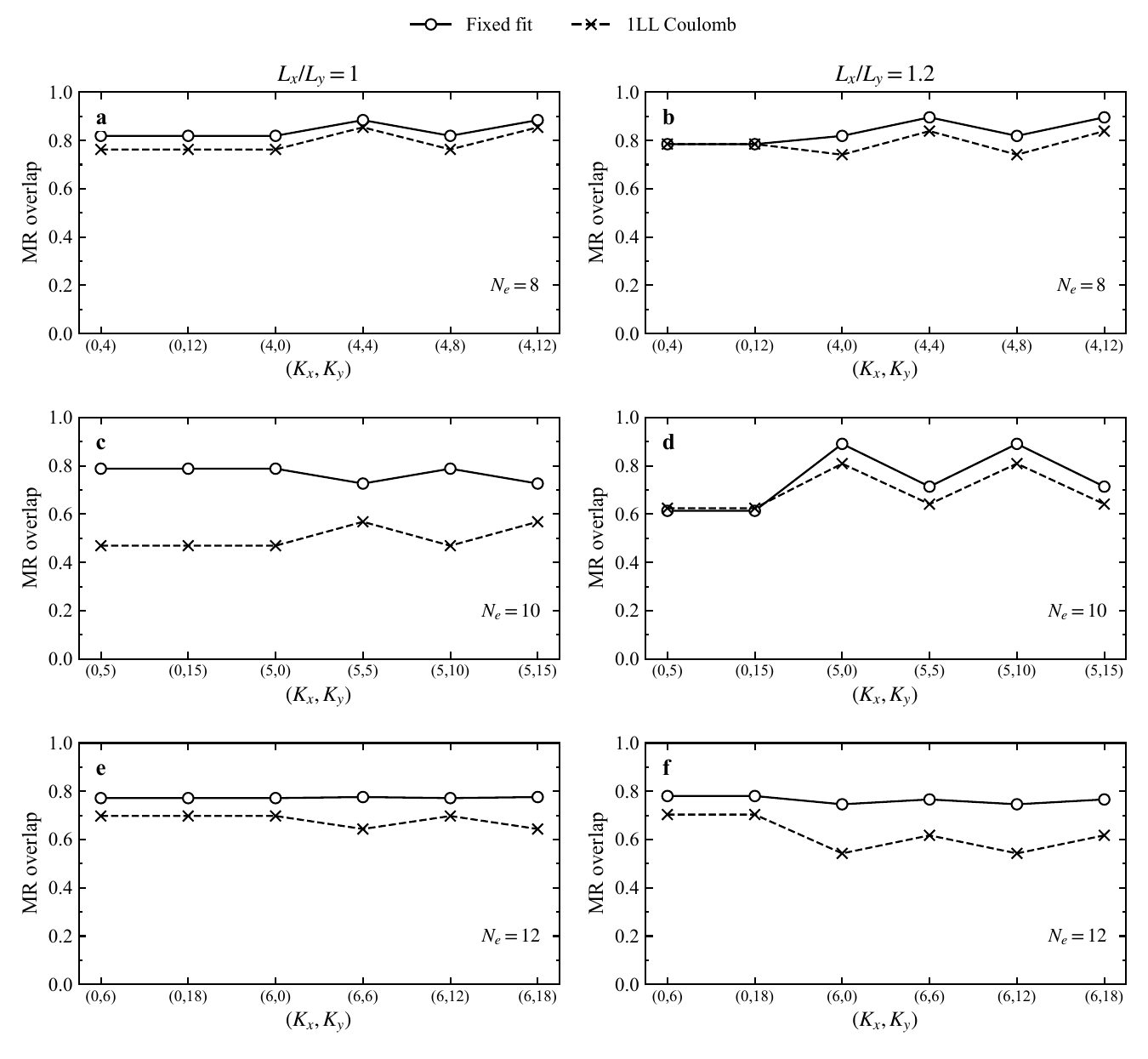}
\caption{Momentum-sector-resolved Moore--Read overlaps on the torus. Rows correspond to $N_e=8$ (\textbf{a,b}), $10$ (\textbf{c,d}), and $12$ (\textbf{e,f}); columns correspond to aspect ratios $L_x/L_y=1$ and $1.2$. Black open circles denote the fixed fitted interaction, and black crosses denote the 1LL Coulomb interaction. Each value is the overlap magnitude between the lowest two-body eigenstate and the zero-energy three-body MR reference state in the same momentum sector, with both states normalized. Connecting lines are guides to the eye. A substantial sector-resolved overlap does not by itself imply that the selected states form an isolated six-state manifold.}
\label{fig:torus_overlaps}
\end{figure}

Supplementary Figures~\ref{fig:torus_square} and~\ref{fig:torus_rectangular} show the low-energy spectra, and Supplementary Figure~\ref{fig:torus_overlaps} compares the overlaps. For $N_e=8$ and $12$, the six selected states are the global lowest six states for both two-body interactions at both aspect ratios. However, their internal splitting and separation from the remaining spectrum depend on size and geometry. For $N_e=10$, states from other sectors intervene within the energy range of the six selected states for both interactions at both aspect ratios, so these states do not form an isolated six-state manifold.

At $N_e=12$ on the square torus, the fitted interaction gives overlaps of approximately $0.7715$--$0.7759$, compared with $0.6436$--$0.6974$ for the 1LL Coulomb interaction. Nevertheless, its six-state internal splitting is larger, $0.01035E_C$ versus $0.00154E_C$, and its separation from the remaining spectrum is smaller, $0.00238E_C$ versus $0.00500E_C$. Here, the internal splitting is the difference between the highest and lowest energies among the six selected states, and the separation is the lowest energy outside this set minus its highest energy. The rectangular $N_e=12$ system likewise shows higher overlaps for the fitted interaction but less favorable spectral isolation and internal splitting than the Coulomb reference.

These results distinguish wave-function overlap from the spectral evidence for a nearly degenerate MR manifold. The fitted interaction retains overlap with the MR reference states across geometries, but does not show a systematic spectral advantage over 1LL Coulomb. The comparison identifies the size and geometry dependence of the transferred interaction and motivates the direct torus optimization below.

\begin{figure}[H]
\centering
\includegraphics[width=\textwidth]{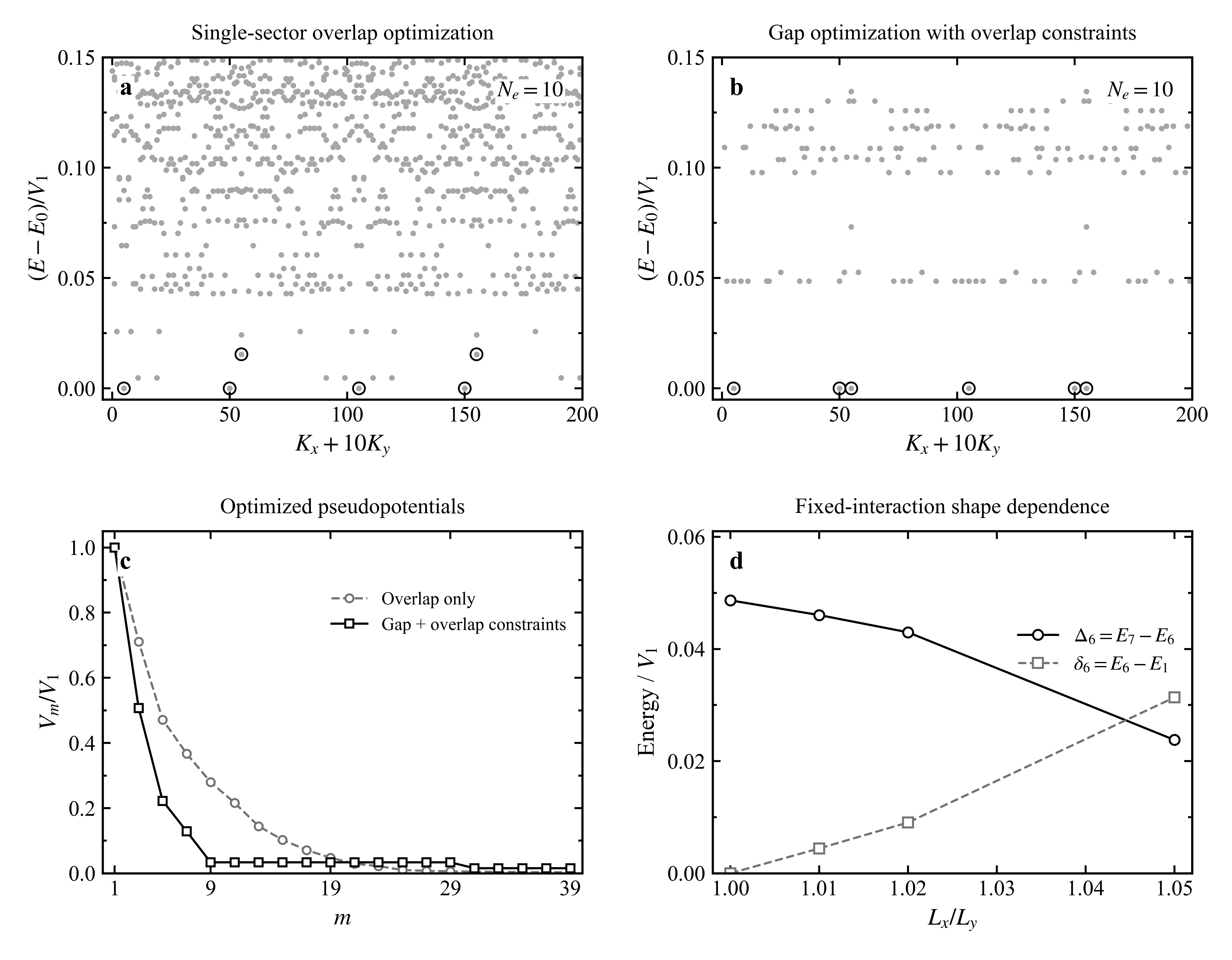}
\caption{Exploratory two-body interaction optimization on the torus. The system contains $N_e=10$ electrons in $N_\phi=20$ orbitals. Twenty odd pseudopotential channels through $m=39$ are retained; $V_1$ is fixed and the remaining nineteen coefficients are optimized subject to nonnegativity and monotonicity. \textbf{a}, Square-torus spectrum obtained by optimizing only the MR overlap in sector $(5,0)$. \textbf{b}, Square-torus spectrum obtained by maximizing the separation from competing states subject to overlap and internal-splitting constraints specified in the text. Grey points show up to eight lowest levels per momentum sector within the displayed window; black open circles mark the lowest states in the six MR parent-state sectors. The horizontal coordinate is $K_x+10K_y$, and excitation energies are measured relative to the respective global ground-state energy $E_0$ in units of $V_1$. \textbf{c}, Pseudopotentials corresponding to \textbf{a} (grey circles) and \textbf{b} (black squares). \textbf{d}, External gap $\Delta_6=E_7-E_6$ (black circles) and internal splitting $\delta_6=E_6-E_1$ (grey squares) for the interaction in \textbf{b}, held fixed as the aspect ratio varies at constant area. Here, $E_1,\ldots,E_7$ are globally ordered energies. Connecting lines are guides to the eye.}
\label{fig:torus_optimize}
\end{figure}

\clearpage
\paragraph*{Exploratory optimization directly on the torus.}
To examine geometry-specific inverse design, we additionally optimize the two-body interaction directly on a square torus with $N_e=10$ and $N_\phi=20$. We retain the twenty odd pseudopotentials $V_1,V_3,\ldots,V_{39}$, set higher channels to zero, and fix $V_1$ as the energy unit while imposing $V_1\ge V_3\ge\cdots\ge V_{39}\ge0$. Both searches start from the normalized fitted profile $V_m/V_1=e^{-0.176(m-1)}$ and use SLSQP. Minimizing $1-|\langle\Psi_{\mathrm{MR}}|\Psi_0\rangle|$ in the $(K_x,K_y)=(5,0)$ sector yields an overlap of approximately $0.8146$, but the selected six sector ground states do not form an isolated lowest-energy manifold [Supplementary Figure~\ref{fig:torus_optimize}\textbf{a}]. At this optimum, the two lowest levels within the $(K_x,K_y)=(5,0)$ sector are nearly degenerate, with an energy separation of approximately $3.0\times10^{-8}V_1$.

We therefore also test a spectral objective with overlap constraints. Introducing an auxiliary energy $g$, we minimize $-g$ subject to $E_b-E_a\ge g$ for every selected target state $a$ and every competing state $b$, and require the internal target-state splitting to be at most $0.2g$. The overlap magnitudes in the two inequivalent target sectors $(0,5)$ and $(5,5)$ are constrained to be at least $0.78$ and $0.72$, respectively; their symmetry partners supply the remaining four states. These thresholds are chosen for this exploratory test. Target-accessible invariant subspaces are tracked during optimization, with their orthogonal complements included among the competing states. Overlap derivatives are evaluated by automatic differentiation and energy derivatives by the Hellmann--Feynman relation. A subsequent scan of all 200 momentum sectors confirms that the resulting six selected states are the global lowest six, with internal splitting $\delta_6=E_6-E_1\simeq2.7\times10^{-12}V_1$ and external gap $\Delta_6=E_7-E_6\simeq0.04866V_1$, where $E_1\le E_2\le\cdots$ are the globally ordered energies. Their MR overlaps are approximately $0.7977$ in $(0,5),(0,15),(5,0),(5,10)$ and $0.7263$ in $(5,5),(5,15)$ [Supplementary Figure~\ref{fig:torus_optimize}\textbf{b,c}].

Keeping this interaction and the area fixed, we then vary $L_x/L_y$ without reoptimization [Supplementary Figure~\ref{fig:torus_optimize}\textbf{d}]. At the sampled ratios $1.01$, $1.02$, and $1.05$, the six selected states remain the lowest six, with $(\delta_6,\Delta_6)/V_1$ approximately $(0.00443,0.04605)$, $(0.00906,0.04297)$, and $(0.03136,0.02378)$, respectively. Thus, spectral isolation persists at these sampled shapes, although the internal splitting grows and exceeds the external gap at $L_x/L_y=1.05$. This example illustrates how the inverse-design objective can incorporate both wave-function and spectral requirements. A systematic study across sizes and geometries is a natural next step for direct torus inverse design.

\clearpage

\clearpage
\section*{Supplementary Note 7. Computational benchmark of gradient evaluation}

We compare automatic differentiation with forward finite differences for the spherical $N_e=12$ system containing 22 orbitals. The optimization fixes $V_1$ and varies the ten odd pseudopotentials $V_3,V_5,\ldots,V_{21}$, subject to the same nonnegativity and monotonicity constraints. Both methods use the same JAX dense-diagonalization calculation of the overlap objective, the same initial pseudopotentials, and precomputed Hamiltonian channel matrices. The difference is the evaluation of the objective gradient: automatic differentiation through the eigendecomposition or forward finite differences.

All runs were performed on the same Apple M5 CPU using JAX 0.7.2 and SciPy 1.16.3, with the same thread settings, \texttt{OMP\_NUM\_THREADS=4} and \texttt{OPENBLAS\_NUM\_THREADS=4}. The SLSQP stopping tolerance was \texttt{ftol}$=10^{-8}$, with a maximum of 2000 iterations. Forward finite differences used absolute steps of $10^{-6}$ and $10^{-7}$ in the numerical pseudopotential coefficients expressed in units of $e^2/(\epsilon\ell_B)$. Reported optimization times exclude preparation and warm-up.

Supplementary Table~\ref{tab:gradient_timing} summarizes the results. Automatic differentiation reaches an overlap of $0.99494327$ in $14810.5$ s, compared with $40254.8$ s and $53797.6$ s for the two finite-difference steps, corresponding to speedups of $2.72$ and $3.63$, respectively. The smaller finite-difference step gives a final overlap within $6.9\times10^{-8}$ of the automatic-differentiation result. Each setting was run once. The comparison uses identical stopping criteria rather than exactly matched final accuracy; the optimization trajectories and iteration counts can differ between gradient methods.

\begin{table}[H]
\centering
\caption{Optimization timing for different gradient methods. The spherical system contains $N_e=12$ electrons in 22 orbitals. Both methods use the same forward calculation, initial pseudopotentials, constraints, and SLSQP stopping settings. Entries report the number of optimizer iterations, the final overlap magnitude with the normalized Moore--Read target state, and the optimization wall-clock time excluding preparation and warm-up. A dash indicates that no finite-difference step is used.}
\label{tab:gradient_timing}
\begingroup
\renewcommand{\arraystretch}{1.3}
\setlength{\tabcolsep}{6pt}
\begin{tabular}{lcccc}
\toprule
Gradient method & Finite-difference step & Iterations & Final overlap & Time (s) \\
\midrule
Automatic differentiation & -- & 24 & 0.99494327 & 14810.5 \\
Forward finite differences & $10^{-6}$ & 16 & 0.99493431 & 40254.8 \\
Forward finite differences & $10^{-7}$ & 23 & 0.99494321 & 53797.6 \\
\bottomrule
\end{tabular}
\endgroup
\end{table}

\section*{Supplementary References}
\bibliography{biblio_fqhe.bib}